\documentclass[10pt,leqno,titlepage]{amsart}
\usepackage[foot]{amsaddr}
\usepackage{amsmath}
\usepackage{graphicx,psfrag,epsf}
\usepackage{enumerate}
\usepackage{natbib}
\usepackage{url} %

\usepackage{mathtools, empheq}
\usepackage{amssymb, setspace}
\usepackage{bm, bbm}
\usepackage{graphicx, tabularx, array}
\usepackage[svgnames,table]{xcolor}
\usepackage{enumitem}
\usepackage{multirow}
\usepackage{hyperref}
\usepackage{geometry}
\usepackage{makecell}
\usepackage{booktabs}
\usepackage{amsmath}
\usepackage{array}
\usepackage{algorithm}
\usepackage{algorithmic}
\usepackage{tikz}
\usepackage{caption}
\usepackage{float}
\usepackage{rotating} 
\usepackage{pdflscape}

\usepackage[normalem]{ulem}

\usepackage{longtable}
\usepackage{threeparttable} %

\usepackage{colortbl} %

\newcolumntype{C}[1]{>{\centering\arraybackslash}m{#1}}

\numberwithin{equation}{section}

\usepackage{amssymb,amsthm,amsmath}

\usepackage{etoolbox}
\AtBeginDocument{%
    \let\orignewpage\newpage 
    \renewcommand\newpage{}
    \patchcmd{\clearpage}{\newpage}{\orignewpage}{}{}}

\begin{document}
\bibliographystyle{plainnat}

\def\spacingset#1{\renewcommand{\baselinestretch}%
{#1}\small\normalsize} \spacingset{1}

\newlist{steps}{enumerate}{1}
\setlist[steps, 1]{label = Step \arabic*:}

\title[Replicable higher-risk combinations of ACEs]{Beyond the ACE Score: Replicable Combinations of Adverse Childhood Experiences That Worsen Depression Risk} 

\author{Ruizhe Zhang$^{1}$}
\author{Jooyoung Kong$^{2}$} 
\author{Dylan S. Small$^{3}$}
\author{William Bekerman$^{3}$}

\thanks{\textbf{Corresponding authors}:\\  William Bekerman, bekerman@wharton.upenn.edu; Ruizhe Zhang, zhangrz22@m.fudan.edu.cn}

\dedicatory{$^{1}$Fudan University, Shanghai, China\\$^{2}$Sandra Rosenbaum School of Social Work, University of Wisconsin-Madison, Madison, WI, USA\\$^{3}$Department of Statistics and Data Science, University of Pennsylvania, Philadelphia, PA, USA}

\begin{abstract}
Adverse childhood experiences (ACEs) are categories of childhood abuse, neglect, and household dysfunction. Screening by a single additive ACE score (e.g., a $\ge 4$ cutoff) has poor individual-level discrimination. We instead identify replicable combinations of ACEs that elevate adult depression risk. Our data turnover framework enables a single research team to explore, confirm, and replicate within one observational dataset while controlling the family-wise error rate. We integrate isotonic subgroup selection (ISS) to estimate a higher-risk subgroup under a monotonicity assumption- additional ACE exposure or higher intensity cannot reduce depression risk. We pre-specify a risk threshold $\tau$ corresponding to roughly a two-fold increase in the odds of depression relative to the no-ACE baseline. Within data turnover, the prespecified component improves power while maintaining FWER control, as demonstrated in simulations. Guided by EDA, we adopt frequency coding for ACE items, retaining intensity information that reduces false positives relative to binary or score codings. The result is a replicable, pattern-based higher-risk subgroup. On held-out BRFSS 2022, we show that, at the same level of specificity (0.95), using our replicable subgroup as the screening rule increases sensitivity by 26\% compared with an ACE-score cutoff, yielding concrete triggers that are straightforward to implement and help target scarce clinical screening resources toward truly higher-risk profiles.
\end{abstract}

\keywords{adverse childhood experiences, sample splitting, exploratory data analysis, replication, subgroup selection}

\maketitle

\spacingset{1.5}

\section{Introduction}

Adverse childhood experiences (ACEs) describe specific categories of early life hardship or harm that occur before an individual reaches adulthood. They include abuse (physical, emotional, or sexual), neglect (physical or emotional), and household challenges such as mental illness, substance use, intimate partner violence, parental separation or divorce, and incarceration. The landmark study conducted by the Centers for Disease Control and Prevention (CDC) and Kaiser Permanente shows that ACEs are common and that they have a clear dose--response relation with multiple negative adult health and social outcomes \citep{felitti1998relationship,dube2003impact,anda2006enduring}. Also, exposure to individual ACEs has been repeatedly associated with a range of adverse physical and emotional health consequences later in life. For example, emotional abuse relates to depression, parental alcohol misuse relates to depression, and maltreatment relates to obesity \citep{chapman2004adverse,anda2002adverse,williamson2002body}. Unfortunately, ACEs often occur together \citep{dong2004interrelatedness,felitti2002relationship}. Large studies show cumulative associations. Each extra category increases the chance of adverse outcomes, and higher counts of ACEs are related to higher risks in many domains \citep{felitti1998relationship,chapman2004adverse,hughes2017effect,merrick2019vital}, which led to the additive 'ACE score'. 

However, the use of a simple additive ACE score is open to question. A single additive score reduces different experiences to one count and treats all events as equal, which is problematic on its face. Two people can share the same score but have very different childhood histories. For example, repeated sexual abuse and parental divorce each add one point. This confuses which exposure patterns should trigger follow-up. Also, the additive ACE score overlooks potential interactions between different ACEs. In clinical and public-health screening and triage, such thresholds give limited individual-level discrimination and can encourage oversimplified or stigmatizing stories \citep{lacey2020practitioner,baldwin2021population,austin2024screening}. Empirical work suggests that although higher ACE counts correlate with worse outcomes at the population level, the score performs poorly for discriminating individuals at risk (e.g., \citealt{meehan2022poor}). In particular, screening based on an ACE threshold (e.g., $\ge4$) may yield high specificity but very low sensitivity (i.e., many at-risk individuals are missed) and only limited improvement in post-test risk. For screening, clinicians and program leads need triggers that are both interpretable and usable. Concrete exposure patterns can guide assessment and services while a single count cannot.

Thus, motivated by this, we focus on combinations of specific ACE items instead of counts. We study adult depression as the primary outcome because it is common and burdensome. Our goal is to identify a higher–risk set of combinations defined by exceeding a prespecified risk threshold—for example, when the odds are at least twice those of people with no ACEs. This yields a screening rule rather than a single score. The rule says which exposure patterns cross a fixed bar. Clinicians can raise or lower the bar to fit the setting. The rule is interpretable and concrete as it points to specific exposure patterns, encodes the simple clinical principle that additional ACEs cannot lower risk of depression, and undergoes a replicability check through our design. Requiring the same high-risk patterns to reappear in distinct subpopulations reduces the chance that an apparent trigger is driven by unmeasured confounding. These features increase causal credibility and practical use of the findings.

We employ a novel statistical design called data turnover (Bekerman et al., in preparation; see \citealp{bekerman2024protocol} for a description). The design splits the data into two complementary parts, and integrates qualitative and quantitative insights from exploratory data analysis (EDA) with structured confirmation. EDA lets us inspect the data, check for anomalies that do not reflect natural variation, and refine questions. EDA also helps flag variables that may not measure the intended constructs and can guide improved definitions and hypotheses \citep{bekerman2024planning}. In observational studies like ours, where treatment assignment is not under our direct control, obtaining replicability (consistent results) across groups that may differ in assignment mechanisms strengthens evidence about the effect under study \citep{rosenbaum2001replicating,rosenbaum2015see}. Prior work often separates these goals. Some methods allow exploration without replication \citep{cox1975note,heller2009split}. Other methods enable replication without exploration \citep{zhao2018cross,karmakar2019integrating}. There are also approaches that cover both exploration and replication but that require two independent teams \citep{roy2022protocol}. Data turnover aims to achieve both exploration and replication with a single research group. 

We combine isotonic subgroup selection (ISS) with data turnover to find higher-risk combinations. ISS builds on a reasonable clinical belief that more ACE exposure should not lower the risk of depression. This creates a partial order on combinations. ISS utilizes this structure to share evidence and to control the family-wise error rate (FWER) in large testing problems. The result is an upward-closed set of higher-risk combinations. We summarize this set by its corner (boundary) combinations, which are easy to read and useful for screening and triage.

We study combinations of ACEs and adult depression using the 2023 Behavioral Risk Factor Surveillance System (BRFSS). BRFSS is a national, ongoing telephone survey run by the CDC. In 2023 it sampled more than 400{,}000 adults across the United States. Ten states\footnote{Delaware, Florida, Georgia, Missouri, Nevada, New Jersey, Oregon, Rhode Island, Tennessee, Virginia.} administered the CDC ACE module in that year. More than 50{,}000 respondents have complete ACE records. About 17{,}000 report zero ACEs, and the rest report at least one ACE. This scale allows precise estimates and supports analysis of ACE combinations. We predefine two complementary parts of the data (red and blue) to implement cross-screening exploration and validation, report both global and replicable findings, and control the FWER throughout.

The remainder of the paper is as follows. Section~\ref{method} includes more details about the dataset, data turnover framework and ISS. We analyzes the BRFSS 2023 data in Section~\ref{application}. In Section~\ref{simulation}, we present the simulation results inspired by the application. Section~\ref{comparison} compares different ways of screening on a held-out dataset. Section~\ref{conclusion} concludes.

\section{Methods}
\label{method}

\subsection{Data}

\begin{figure}
  \centering
  \includegraphics[width=\linewidth]{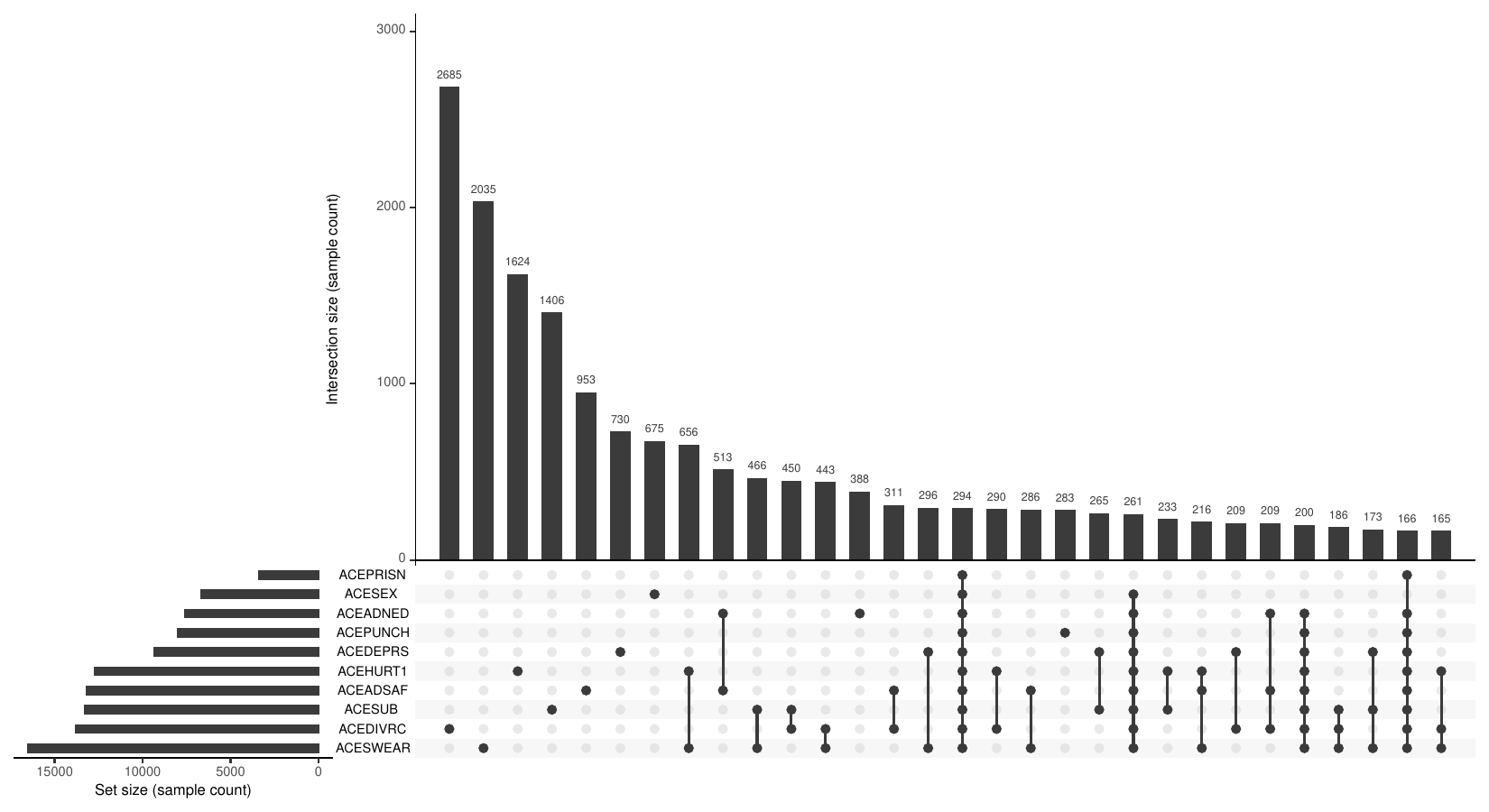}
  \caption{UpSet plot of ACE combinations. 
  Bars at the top show intersection sizes (sample counts). 
  Connected filled dots mark which ACE items co-occur in each intersection. 
  Horizontal bars on the left show marginal set sizes for each item (sample counts). 
  The figure displays the largest intersections in the BRFSS\,2023 analytic sample and provides a compact view of both prevalence and co-occurrence.}
  \label{fig:ace_upset}
\end{figure}

We analyze the 2023 BRFSS ACE module from ten states (listed above), with $>$50{,}000 complete ACE records. Depression is coded $Y{=}1$ if the answer to \emph{'(Ever told) (you had) a depressive disorder (including depression, major depression, dysthymia, or minor depression)?'} is \emph{'Yes'} and $0$ otherwise. Following previous studies utilizing the BRFSS data \citep{bhan2014childhood,gilbert2015childhood,waehrer2020disease}, we collapse different questions into particular indicators. More details are provided in table \ref{tab:data_processing}. In total, we create 10 binary ACE items from the BRFSS data (see table \ref{tab:ace_items_brfss}), which is consistent with the categorization in CDC-Kaiser study \citep{felitti1998relationship}. Figure~\ref{fig:ace_upset} summarizes marginal prevalence and co-occurrence of ACE items in BRFSS 2023. We display only the top 30 combinations by size. The plot mainly shows two messages. First, item prevalence is uneven across different ACEs. Second, co-occurrence is common. Many respondents report multi-item exposure rather than a single item. For example, the pair \emph{ACEHURT1 \& ACESWEAR} is among the largest intersections (rank 8 of 30) with 656 respondents, and even the set with all the 10 ACEs appears (rank 16 of 30) with 294 respondents. These patterns are consistent with our study design that we focus on combinations rather than counts.

To support the data turnover design, we split the BRFSS sample into two parts for exploration, confirmation, and replication.  We form two subpopulations in BRFSS 2023. One includes respondents in states that voted for the Democratic candidate in the 2024 U.S. presidential election ('blue states'). The other includes respondents in states that voted for the Republican candidate ('red states'). These two sets of states often differ in social, economic, and health policy. Prior work shows that socio-political context relates to childhood maltreatment \citep{ulke2021socio}. For example, states may adopt different strategies toward public health programs and in the scope of social safety nets. Such differences may affect the prevalence of ACEs and the mechanisms of exposure. They may also lead to different patterns of co-occurrence across ACE items.

After processing the data, we restrict our sample to those who do not have any missing values in any of the ten ACE variables and depression measurement. In total, there are 49,547 remaining individuals, among whom 23,390 individuals are from blue states\footnote{Delaware, New Jersey, Oregon, Rhode Island, Virginia} and 26,157 individuals from red states\footnote{Florida, Georgia, Missouri, Nevada, Tennessee}. In the part of data with individuals from blue states, 7,363 individuals do not have any ACEs while 16,027 were exposed to at least one ACE. In the part of data with individuals from red states, 8,479 individuals do not have any ACEs while 17,678 were exposed to at least one ACE.

\begin{table}
\small\sf\centering
\caption{Data Processing Steps and Collapsed Indicators}
\label{tab:data_processing}
\begin{tabular}{@{}p{.2\textwidth} p{.45\textwidth} p{.3\textwidth}@{}}
\hline
\textbf{ACE Items} & \textbf{Questions Collapsed} & \textbf{Encoding Method} \\ \hline
Any Sexual Abuse & Number of times forced to touch, be touched, or have sex with anyone at least five years older or an adult & Constructed by summing responses across multiple questions: all “never” encoded as 0, otherwise encoded as 1  \\ \hline
Household Substance Use & Alcohol and illegal drug use in the household & Constructed by summing responses across multiple questions: all “never” encoded as 0, otherwise encoded as 1 \\ \hline
Physical Abuse & Frequency of physical abuse & ‘Once’ or ‘More than once' encoded as 1, ‘None’ encoded as 0 \\ \hline
Verbal Abuse & Frequency of verbal abuse & ‘Once’ or ‘More than once' encoded as 1, ‘None’ encoded as 0 \\ \hline
Parental Divorce & Response to parental divorce question & 'Yes' encoded as 1, 'No' encoded as 0 \\ \hline
Living with a Depressed or Mentally Ill Person & Response to question about living with a mentally ill person & 'Yes' encoded as 1, 'No' encoded as 0 \\ \hline
Living with an Incarcerated Person & Response to question about living with an incarcerated person & 'Yes' encoded as 1, 'No' encoded as 0 \\ \hline
Physical Violence Between Parents & Frequency of physical violence between parents & ‘Once’ or ‘More than once' encoded as 1, ‘None’ encoded as 0 \\ \hline
Emotional Neglect & For how much of your childhood was there an adult in your household who made you feel safe and protected & ‘Never’ and ‘A little of the time’ encoded as 1, ‘Some of the time’, ‘Most of the time’, ‘All of the time’ encoded as 0 \\ \hline
Physical Neglect & For how much of your childhood was there an adult in your household who tried hard to make sure your basic needs were met & ‘Never’ and ‘A little of the time’ encoded as 0, ‘Some of the time’, ‘Most of the time’, ‘All of the time’ encoded as 1 \\ \hline
\end{tabular}
\end{table}

\begin{table}
\small\sf\centering
\caption{ACE Items in BRFSS 2023}
\label{tab:ace_items_brfss}
\begin{tabular}{ll}
  \hline
  \textbf{ACE Item} & \textbf{Variable} \\
  \hline
  Live With Anyone Depressed, Mentally Ill, Or Suicidal & ACEDEPRS \\
  Substance Abuse in the Home & ACESUB \\
  Live With Anyone Who Served Time in Prison or Jail & ACEPRIS \\
  Were Your Parents Divorced/Separated & ACEDIVRC \\
  How Often Did Your Parents Beat Each Other Up & ACEPUNCH \\
  How Often Did A Parent Physically Hurt You In Any Way & ACEHURT1 \\
  How Often Did A Parent Swear At You & ACESWEAR \\
  Sexual Abuse & ACESEX \\
  Did An Adult Make You Feel Safe And Protected & ACEADSAF \\
  Did An Adult Make Sure Basic Needs Were Met & ACEADNED \\
  \hline
\end{tabular}
\end{table}

\subsection{Isotonic Subgroup Selection}

We use isotonic subgroup selection (ISS) to identify the high-risk group \citep{muller2025isotonic}. Let $(X_i,Y_i)$ be independent for $i=1,\dots,n$. The vector $X_i\in\mathbb{R}^d$ contains covariates. The variable $Y_i$ is the outcome. In this study, $Y_i=1$ indicates depression and $Y_i=0$ otherwise. The vector $X_i$ encodes combinations of ten binary ACE indicators.

We define $\eta(x)=\mathbb{E}(Y\mid X=x)$ and assume coordinate-wise monotonicity: if $x\preceq x'$, then $\eta(x)\le \eta(x')$. The relation $x\preceq x'$ means $x_j\le x'_j$ for all $j=1,\dots,d$. This assumption encodes a simple idea: added ACE exposure should not reduce the probability of depression. For example, a profile with physical abuse and parental substance use should have risk at least as high as a profile with only parental substance use.

Fix a threshold $\tau$. The target set is the superlevel region
\[
X_\tau(\eta)=\{\,x:\ \eta(x)\ge\tau\,\},
\]
interpreted as the subgroup of profiles whose depression probability exceeds $\tau$. ISS estimates a subset $\hat A$ that approximates $X_\tau(\eta)$ with principled error control.

\paragraph{Step 1: hypothesis construction.}
For each observed profile $X_i$, test
\[
H_{0,i}:\ \eta(X_i)<\tau
\quad\text{vs.}\quad
H_{1,i}:\ \eta(X_i)\ge\tau.
\]
Because $Y$ is binary, we use anytime-valid martingale $p$-values for bounded outcomes. The construction aggregates evidence from profiles $X_j\preceq X_i$, comparing the observed number of depressed cases to what would be expected under $\eta(X_i)<\tau$. This yields valid $p_i$ even under dependence across tests.

Formally, let $I(x)=\{\,j:\ X_j\preceq x\,\}$ and $n(x)=|I(x)|$. Write the associated responses as $Y_{(1)}(x),\dots,Y_{(n(x))}(x)$, and set $S_k(x)=\sum_{j=1}^k Y_{(j)}(x)$. An anytime-valid $p$-value for testing $\eta(x)<\tau$ is
\[
p_\tau(x)=\min_{1\le k\le n(x)}
\frac{\tau^{S_k(x)}(1-\tau)^{\,k-S_k(x)+1}}
     {B\!\bigl(1-\tau;\,k-S_k(x)+1,\,S_k(x)+1\bigr)},
\]
where $B(z;a,b)$ is the incomplete Beta function. Then
$\Pr\!\bigl(p_\tau(x)\le\alpha\mid X\bigr)\le\alpha$ whenever $\eta(x)<\tau$ \citep[Lemma~22]{muller2025isotonic}.

\paragraph{Step 2: logical structure and multiple testing.}
Let $\{x^{(1)},\dots,x^{(m)}\}$ be the distinct observed profiles and index the corresponding hypotheses by $i=1,\dots,m$. The nulls are logically related: if $x^{(i)}\preceq x^{(j)}$, then $H_{0,j}\subseteq H_{0,i}$. Encode these relations in a directed acyclic graph (DAG) $G=(I,E)$ with $I=\{1,\dots,m\}$ and an edge $i\to j$ whenever $x^{(j)}\preceq x^{(i)}$ (edges point from stronger to weaker profiles). Form a polyforest $F$ by keeping at most one parent per node. The parent choice is not unique; one may use the closest parent in $\ell_\infty$ norm or another distance.

Let $R^{\text{ISS}}_\alpha(G,p)$ be the rejection set from the DAG algorithm. Distribute the global level $\alpha$ across the roots of $F$; each root $i$ receives
\[
\alpha_i=\frac{\#\{\text{leaf descendants of } i \text{ in } F\}}
               {\#\{\text{all leaf nodes in } F\}}\cdot\alpha.
\]
Reject a root $i$ if $p_{\tau,i}\le\alpha_i$, and then reject all ancestors of $i$ in the full DAG $G$. Remove rejected nodes from $F$, reallocate unused $\alpha$ to the remaining roots, and iterate until no further rejections occur \citep[Algorithm~1]{muller2025isotonic}. The final rejection set respects the partial order and controls the family-wise error rate (FWER).

\paragraph{Step 3: upward closure.}
Let $R^{\text{ISS}}_\alpha(G,p)$ be the set of rejected nodes. Define
\[
\hat A=\{\,x:\ x\succeq x^{(i)}\ \text{for some } i\in R^{\text{ISS}}_\alpha(G,p)\,\}.
\]
Thus $\hat A$ is the upward closure of $R^{\text{ISS}}_\alpha(G,p)$ under $\preceq$. If a profile $x^{(i)}$ is high-risk, every more severe profile $x\succeq x^{(i)}$ is included.

Under this construction, ISS controls the FWER: with probability at least $1-\alpha$, no profile with $\eta(x)<\tau$ enters $\hat A$. More formally, conditional on the observed $X$, ISS returns a data–dependent set $\hat A$ such that  $\mathbb{P}\bigl(\hat A \subseteq X_\tau(\eta)\,\big|\,X\bigr)\;\ge\;1-\alpha$, uniformly over the monotone model class considered in \citet[Theorem~9]{muller2025isotonic}. Equivalently, $\mathbb{P}\bigl(\exists\,x\in\hat A:\ \eta(x)<\tau \,\big|\,X\bigr)\;\le\;\alpha$. ISS provides a conservative and rigorous way to define ACE exposure patterns linked to higher depression risk. The selected subgroups include only ACE combinations whose estimated depression probability exceeds the threshold. The procedure uses valid error control and the monotonicity structure to yield interpretable high-risk subgroups.

\subsection{Data Turnover: Design and Implementation}

We employ data turnover \citep{bekerman2024protocol} to analyze the data, which combines screening with data-informed exploration while keeping strong family-wise error control. The approach is more flexible than automated cross-screening \citep{zhao2018cross} as it allows exploratory data analysis (EDA) in one part of the sample and still guarantees valid inference \citep[e.g.,][]{bekerman2024planning}. We split the sample into a red part and a blue part. The roles are fixed before any analysis. The red part follows a pre-specified screening and validation plan \citep{zhang2025protocol} without EDA. The blue part permits EDA. Both parts are used to screen and to validate, but only the blue part can generate novel hypotheses for the red part. In other words, when we formulate the data analysis plan on the red part, we may use extra information from the blue part rather than strictly follow a pre-specified one. The whole procedure strictly controls the family-wise error rate (FWER). We call a result replicable if it shows a significant effect in both groups \citep{bogomolov2023replicability}. We also test a global null hypothesis expressed, i.e., identifying outcomes that are affected in at least one of the two subpopulations. 

\paragraph{Threshold and target.}
For each observed item-set $X_i$, we test
\[
H_{0,i}:\ \eta(X_i)<\tau \quad\text{vs.}\quad H_{1,i}:\ \eta(X_i)\ge\tau,
\]
where $\eta(x)=\Pr(Y=1\mid X=x)$ and $Y$ indicates depression. As stated in the protocol, we pre-specify $\tau=0.172$, which is interpreted as at least double the odds of depression compared with the group with no ACEs.

\paragraph{Step 1: Screening in the red part (no EDA).}
Compute ISS $p$-values $p^{R}_{\tau,i}$ for combinations of ACEs in the red part. Form the screened set $\mathcal{S}_\kappa \;=\; \bigl\{\, i:\ p^{R}_{\tau,i}\le \kappa \,\bigr\}$, with the pre-specified cutoff $\kappa = \alpha/2 = 0.025$. Build a polyforest on $\mathcal{S}_\kappa$ to encode the monotone logical order used by ISS. Let $G$ be the full DAG over candidate combinations. For $i\in\mathcal{S}_\kappa$, define the cover set $\mathrm{Cover}(i)\;=\;\bigl\{\, j\in\mathcal{S}_\kappa:\ X_j\succ X_i \ \text{and there is no } k\in\mathcal{S}_\kappa \text{ with } X_j\succ X_k\succ X_i \,\bigr\}$.
Select at most one parent via red-side evidence: $\mathrm{par}(i)\ \in\ \arg\min_{j\in \mathrm{Cover}(i)}\ p^{R}_{\tau,j}$, breaking ties at random. If $\mathrm{Cover}(i)=\varnothing$, declare $i$ a root. Denote the resulting structure by $F^{B}_\kappa$. This routes $\alpha$ toward chains with stronger upstream evidence and reduces $\alpha$–blocking by weak parents. Further explanations and illustrative examples are provided in Appendix~\ref{app:evidence_guided_details}. Send $(\mathcal{S}_\kappa, F^{B}_\kappa)$ to the blue part for validation.

\paragraph{Step 2: EDA on the blue part.}
First, we could repeat the pre-specified screening steps in the blue part. We use the blue data to select the candidate combinations to test in the red part. We also use blue-side evidence to set the polyforest structure for red validation. Notably, we perform EDA only in the blue part and then prepare a validation plan for the red part.

\paragraph{Step 3: Two-way validation and error control.}
Validate the red-screened set $\mathcal{S}_\kappa$ in the blue part by running ISS on $F^{B}_\kappa$ at a pre-assigned level $\alpha_B$ (we use $\alpha_B=\alpha/2$). Validate the blue-generated hypotheses in the red part at level $\alpha_R$ (we use $\alpha_R=\alpha/2$). The split satisfies $\alpha_B+\alpha_R\le \alpha$ and yields strong FWER control for all claims combined. We report two types of findings. A combination is replicable if both parts reject the null hypothesis that the probability of depression conditional on this combination exceeds the threshold. The global null for a combination is rejected if at least one part rejects the null hypothesis.

ISS supplies anytime-valid $p$-values under monotonicity and controls multiplicity over the structured family through its DAG procedure \citep{muller2025isotonic}. Data turnover decides which hypotheses enter validation, how parents are chosen for the polyforest, and provides more flexibility through EDA. More details about the pre-specified plan is in the protocol \citep{zhang2025protocol}. 

\section{Application}
\label{application}

We analyze BRFSS 2023 with the ACE module. The outcome is a binary indicator for depression. We report results in both directions - Red$\rightarrow$Blue and Blue$\rightarrow$Red. Specifically, Red$\rightarrow$Blue strictly follow the prespecified plan \citep{zhang2025protocol}. As for Blue$\rightarrow$Red, we do EDA on the blue part to formulate the data analysis plan on the red part. After completing both data analysis, we report two sets of findings. First, replicable profiles are those combinations rejected in both the red and blue parts (each tested at level $\alpha/2$). Second, global discoveries are profiles for which the global null $H_{0,i}^{\text{red}}\cap H_{0,i}^{\text{blue}}$ is rejected, i.e., at least one part rejects its corresponding null. 

Since the higher-risk group may include many profiles, we report the corner combinations. They are the boundary points of the upward-closed region. A corner combination is a minimal selected profile under the coordinate-wise order: it is in the set, but if you remove any ACE from it, the resulting profile is not. Reporting corners is sufficient because, by monotonicity, any profile that adds ACEs to a corner is also included.

\subsection{Red$\rightarrow$Blue: Pre-specified screening and validation}

We follow the pre-specified plan. We compute ISS $p$–values in the red part. We screen with the cutoff $\kappa=\alpha/2=0.025$ and obtain the candidate set $\mathcal{S}_\kappa=\{i:\,p_{\tau,i}^{\text{red}}\le 0.025\}$. We then validate these candidates in the blue part. We build a polyforest on $\mathcal{S}_\kappa$ using the evidence–guided rule. We run the ISS DAG test at level $\alpha/2=0.025$ and take the upward closure. This yields the selected set $\widehat A_{R\to B}^{\text{bin}}$.

The blue–side validation rejects $77$ candidates before closure and selects $2$ corner combinations after upward closure: $\bigl\{
\{\texttt{ACEDEPRS}\},
\;\{\texttt{ACESUB},\ \texttt{ACEDIVRC},\ \texttt{ACESWEAR},\ \texttt{ACESEX}\}
\bigr\}$. By monotonicity, every profile that adds any further ACEs to either corner is included. On the binary grid ($2^{10}$ profiles), $\widehat A_{R\to B}^{\text{bin}}$ covers $544/1024=53.1\%$ of profiles. 
In the BRFSS 2023 sample, it flags $9{,}842$ of $49{,}547$ respondents ($19.9\%$) under this rule.

\subsection{Blue$\rightarrow$Red: EDA}

In this section, we analyzed the data on the blue subgroup, the one with a smaller sample size, to plan the analysis
for the red data. Below we provide a
detailed description on how we conduct the exploratory data analysis (EDA) and prepared the plan for analysis on the red subgroup.

Prior BRFSS studies collapse multiple questionnaire items into binary indicators for each ACE domain. Yet several BRFSS ACE questions offer frequency choices such as \emph{None}, \emph{Once}, and \emph{More than once}. We therefore try to bring in frequency information. Concretely, we still collapse multiple questions into the ten standard ACE domains, but for four items (\texttt{ACEPUNCH}, \texttt{ACEHURT1}, \texttt{ACESWEAR}, \texttt{ACESEX}) we code \emph{0 = never}, \emph{1 = once}, \emph{2 = $\ge$2 times}. In addition, two items (\texttt{ACEADSAF}, \texttt{ACEADNED}) report five categories: \emph{Never}, \emph{A little of the time}, \emph{Some of the time}, \emph{Most of the time}, \emph{All of the time}. Because these questions are framed in protective terms, we reverse code them so that larger values indicate worse adversity: \emph{All of the time}$\to 0$, \emph{Most}$\to 1$, \emph{Some}$\to 2$, \emph{A little}$\to 3$, \emph{Never}$\to 4$. More details appear in Table~\ref{tab:freq_data_processing}. Note that under frequency coding, the monotonicity assumption implies that increasing exposure, either by adding ACE items or by raising the frequency level of any item, cannot lower the probability of depression (i.e., if $x\preceq x'$ coordinate–wise, then $\eta(x')\ge \eta(x)$).

\begin{table}
\caption{Data Processing Steps and Collapsed ACE Items (Frequency Coding)}
\small\sf\centering
\begin{tabular}{@{}p{.1\textwidth} p{.4\textwidth} p{.4\textwidth}@{}}
\hline
\textbf{ACE Item} & \textbf{Questions Collapsed} & \textbf{Encoding Method} \\ \hline
ACESEX & Number of times forced to touch, be touched, or have sex with anyone at least five years older or an adult & Constructed by summing responses across multiple questions: all “never” encoded as 0, total frequency $=1$ encoded as 1, total frequency $\geq2$ encoded as 2  \\ \hline
ACESUB & Alcohol and illegal drug use in the household & Constructed by summing responses across multiple questions: all “never” encoded as 0, otherwise encoded as 1 \\ \hline
ACEHURT1 & Frequency of physical abuse & ‘More than once' encoded as 2, 'Once' encoded as 1 and ‘None’ encoded as 0 \\ \hline
ACESWEAR & Frequency of verbal abuse & ‘More than once' encoded as 2, 'Once' encoded as 1 and ‘None’ encoded as 0 \\ \hline
ACEDIVRC & Response to parental divorce question & 'Yes' encoded as 1, 'No' encoded as 0 \\ \hline
ACEDEPRS & Response to question about living with a mentally ill person & 'Yes' encoded as 1, 'No' encoded as 0 \\ \hline
ACEPRISN & Response to question about living with an incarcerated person & 'Yes' encoded as 1, 'No' encoded as 0 \\ \hline
ACEPUNCH & Frequency of physical violence between parents & ‘More than once' encoded as 2, 'Once' encoded as 1 and ‘None’ encoded as 0 \\ \hline
ACEADSAF & For how much of your childhood was there an adult in your household who made you feel safe and protected & Five-level ordered (reverse-coded): All of the time = 0; Most of the time = 1; Some of the time = 2; A little of the time = 3; Never = 4 \\ \hline
ACEADNED & For how much of your childhood was there an adult in your household who tried hard to make sure your basic needs were met & Five-level ordered (reverse-coded): All of the time = 0; Most of the time = 1; Some of the time = 2; A little of the time = 3; Never = 4 \\ \hline
\end{tabular}
\label{tab:freq_data_processing}
\end{table}

We then run ISS on the blue part under two specifications: (a) the traditional binary coding for all ten domains; (b) the frequency coding in which the four items above are three-level categorical and two are five-level ordered after reverse coding. The two analyses yield markedly different pictures of the high–risk exposure structure (Tables~\ref{tab:binary coding}--\ref{tab:frequency coding}).

\begin{landscape}
\begin{table}
\centering
\small\sffamily
\caption{ISS-selected corners on Blue: \emph{binary} coding}
\label{tab:binary coding}
\begin{tabular}{@{}c *{10}{c}@{}}
\toprule
 & ACEDEPRS & ACESUB & ACEPRISN & ACEDIVRC & ACEPUNCH & ACEHURT1 & ACESWEAR & ACESEX & ACEADSAF & ACEADNED \\ 
\midrule\addlinespace[2.5pt]
Corner 1 & 1 & 0 & 0 & 1 & 1 & 1 & 1 & 0 & 0 & 0 \\ 
Corner 2 & 1 & 0 & 0 & 0 & 1 & 1 & 1 & 1 & 0 & 0 \\ 
\bottomrule
\end{tabular}
\end{table}

\begin{table}
\small\sf\centering
\small
\caption{ISS-selected corners on Blue: \emph{frequency} coding}
\begin{tabular}{@{}c *{10}{c}@{}}
\toprule
 & ACEDEPRS & ACESUB & ACEPRISN & ACEDIVRC & ACEPUNCH & ACEHURT1 & ACESWEAR & ACESEX & ACEADSAF & ACEADNED \\ 
\midrule\addlinespace[2.5pt]
Corner 1 & 1 & 0 & 0 & 0 & 0 & 0 & 2 & 2 & 0 & 0 \\ 
Corner 2 & 1 & 1 & 0 & 0 & 0 & 0 & 2 & 0 & 0 & 0 \\ 
Corner 3 & 1 & 0 & 0 & 1 & 0 & 1 & 2 & 1 & 1 & 1 \\ 
Corner 4 & 0 & 0 & 0 & 0 & 0 & 1 & 2 & 2 & 1 & 0 \\ 
Corner 5 & 0 & 1 & 0 & 0 & 0 & 0 & 2 & 2 & 1 & 0 \\ 
Corner 6 & 1 & 1 & 1 & 0 & 0 & 0 & 0 & 2 & 1 & 0 \\ 
Corner 7 & 0 & 0 & 0 & 1 & 0 & 0 & 2 & 2 & 1 & 0 \\ 
Corner 8 & 1 & 0 & 0 & 0 & 0 & 2 & 2 & 0 & 2 & 1 \\ 
Corner 9 & 1 & 0 & 0 & 1 & 0 & 1 & 2 & 0 & 2 & 1 \\ 
Corner 10 & 0 & 1 & 0 & 1 & 0 & 0 & 2 & 2 & 0 & 0 \\ 
Corner 11 & 0 & 0 & 1 & 1 & 0 & 1 & 2 & 2 & 0 & 0 \\ 
Corner 12 & 1 & 1 & 0 & 1 & 1 & 0 & 0 & 2 & 1 & 0 \\ 
Corner 13 & 1 & 1 & 0 & 1 & 0 & 1 & 0 & 2 & 1 & 0 \\ 
Corner 14 & 1 & 0 & 1 & 1 & 1 & 1 & 2 & 1 & 1 & 0 \\ 
Corner 15 & 1 & 0 & 0 & 0 & 0 & 0 & 2 & 0 & 3 & 1 \\ 
Corner 16 & 1 & 0 & 0 & 1 & 0 & 0 & 2 & 1 & 2 & 1 \\ 
Corner 17 & 1 & 1 & 0 & 0 & 0 & 0 & 1 & 2 & 1 & 0 \\ 
Corner 18 & 1 & 0 & 0 & 1 & 2 & 2 & 2 & 1 & 3 & 0 \\ 
Corner 19 & 1 & 1 & 0 & 1 & 0 & 0 & 1 & 2 & 0 & 0 \\ 
Corner 20 & 1 & 0 & 1 & 1 & 0 & 0 & 2 & 0 & 2 & 1 \\ 
Corner 21 & 1 & 1 & 0 & 0 & 0 & 1 & 1 & 2 & 0 & 0 \\ 
Corner 22 & 1 & 0 & 0 & 0 & 0 & 1 & 2 & 1 & 2 & 1 \\ 
\bottomrule
\label{tab:frequency coding}
\end{tabular}
\end{table}
\end{landscape}

Under the binary coding, ISS identifies only two corner combinations (Table~\ref{tab:binary coding}). Both require \texttt{ACEDEPRS}$=1$, \texttt{ACEPUNCH}$=1$, \texttt{ACEHURT1}$=1$, \texttt{ACESWEAR}$=1$, and either \texttt{ACEDIVRC}$=1$ or \texttt{ACESEX}$=1$. When frequency information is incorporated (Table~\ref{tab:frequency coding}), the picture changes substantially. ISS now detects several distinct corners. For example, in Corner~1 of Table~\ref{tab:frequency coding}, only \texttt{ACEDEPRS}$=1$, \texttt{ACESWEAR}$=2$ and \texttt{ACESEX}$=2$  are required. Relative to Corner~2 in Table~\ref{tab:binary coding}, this shows that experiencing verbal abuse more than once, experiencing sexual abuse more than once together with living with a depressed or mentally ill household member, suffices to exceed the risk threshold even without physical abuse or interparental violence.  In addition, in most corners of Table~\ref{tab:frequency coding}, \texttt{ACESEX}$=2$ or \texttt{ACESWEAR}$=2$ are required, indicating that repeated sexual/verbal abuse can push many combinations above the threshold and highlighting the severity of repeated exposure to sexual abuse.

Conclusively, the binary approach implicitly mixes individuals exposed once with those exposed repeatedly, assuming the same level of risk across very different exposure intensities. This masks meaningful differences in the intensity of adverse experiences. The frequency–augmented approach distinguishes one-time from repeated exposures, revealing that repeated verbal and sexual abuse have a disproportionate impact on depression risk. This provides a clearer and more interpretable characterization of high-risk profiles, motivating the use of frequency–augmented coding in the red subgroup analysis.

Then, we mirror the pre-specified plan: use blue to help the red analysis by (i) screening to shrink the set of combinations to be tested and (ii) organizing the polyforest used for testing on the red side, aiming to remove unpromising profiles, reduces multiplicity, and increases power at the same total error level. Throughout this Blue$\rightarrow$Red path we adopt the frequency encoding for both screening and validation. Other implementation details are exactly the same as the pre-specified Red$\rightarrow$Blue plan. 

Guided by blue-side EDA, we considered a tiered $\alpha$ allocation that prioritizes combinations featuring ACEs judged more consequential (e.g., \texttt{ACEDEPRS}, \texttt{ACESEX}). We built marginal risk-ratio summaries and a conditional dominance map to inform the tiers (Figures~\ref{fig:marginal}–\ref{fig:conditional}). We first evaluated tiering in the exploratory blue part. This run did not use candidate screening or evidence‐guided parenting. Adding tiering to ISS produced a more selective set. The tiered set was smaller than the non‐tiered set and mainly pruned combinations, with few new additions. We then ran simulations that used cross‐part evidence for screening and parent selection. Evidence‐guided ISS with and without tiering showed almost the same rejection patterns and discovery counts.  Complete exploration details and results appear in Appendix~\ref{app:tiered allocation} and Figure~\ref{fig:tiering}.

Following this analysis plan informed by EDA on the blue part, we analyze the red part. The screened candidate set contains \(4{,}616\) unique ACE combinations to be tested on the red side. 
This is an \(85.8\%\) reduction relative to the full frequency–coded grid of \(2^4 \times 3^4 \times 5^2 = 32{,}400\) combinations. The red–side validation rejects $3{,}206$ candidates before closure and selects $17$ corner point after upward closure (see Table~\ref{tab:redtoblue} for more details.). By monotonicity under frequency coding, the selected set is upward–closed in both breadth and intensity. Any profile that adds further ACE items (turning a 0 into a positive level) is included, and any profile that raises the frequency level of an already present ACE is also included. On the frequency grid ($32{,}400$ profiles), the upward–closed selection covers $6{,}166$ profiles, i.e., $19.03\%$ of the grid. 
In the BRFSS 2023 sample with frequency coding ($N=49{,}547$), the rule flags $4{,}364$ respondents, i.e., $8.81\%$ of individuals.

\begin{landscape}
    \begin{table}
\small\sf\centering
\small
\caption{Corner Combinations (blue $\to$ red)  under \emph{frequency} coding}
\begin{tabular}{@{}c *{10}{c}@{}}
\toprule
 & ACEDEPRS & ACESUB & ACEPRISN & ACEDIVRC & ACEPUNCH & ACEHURT1 & ACESWEAR & ACESEX & ACEADSAF & ACEADNED \\ 
\midrule\addlinespace[2.5pt]
Corner 1 & 1 & 0 & 0 & 0 & 0 & 0 & 2 & 1 & 0 & 0 \\ 
Corner 2 & 1 & 1 & 1 & 1 & 0 & 0 & 0 & 2 & 0 & 0 \\ 
Corner 3 & 1 & 1 & 0 & 1 & 0 & 1 & 1 & 2 & 0 & 0 \\ 
Corner 4 & 1 & 1 & 0 & 0 & 0 & 0 & 2 & 0 & 1 & 0 \\ 
Corner 5 & 1 & 1 & 1 & 0 & 0 & 0 & 0 & 2 & 1 & 0 \\ 
Corner 6 & 1 & 1 & 0 & 1 & 0 & 0 & 0 & 2 & 1 & 0 \\ 
Corner 7 & 0 & 1 & 0 & 1 & 1 & 1 & 2 & 2 & 1 & 0 \\ 
Corner 8 & 1 & 0 & 0 & 1 & 1 & 1 & 2 & 0 & 2 & 0 \\ 
Corner 9 & 1 & 0 & 1 & 0 & 1 & 2 & 2 & 0 & 2 & 0 \\ 
Corner 10 & 0 & 1 & 0 & 1 & 1 & 0 & 2 & 2 & 2 & 0 \\ 
Corner 11 & 1 & 1 & 1 & 0 & 1 & 1 & 2 & 0 & 0 & 1 \\ 
Corner 12 & 1 & 1 & 0 & 1 & 0 & 1 & 0 & 2 & 0 & 1 \\ 
Corner 13 & 1 & 0 & 1 & 0 & 0 & 1 & 2 & 0 & 1 & 1 \\ 
Corner 14 & 0 & 1 & 0 & 1 & 1 & 0 & 2 & 2 & 1 & 1 \\ 
Corner 15 & 1 & 0 & 0 & 1 & 0 & 1 & 2 & 0 & 2 & 1 \\ 
Corner 16 & 1 & 0 & 0 & 0 & 0 & 2 & 2 & 0 & 2 & 1 \\ 
Corner 17 & 0 & 1 & 0 & 1 & 0 & 2 & 2 & 2 & 2 & 1 \\

\bottomrule
\label{tab:redtoblue}
\end{tabular}
\end{table}
\end{landscape}

\subsection{Replicable Findings and Global Discoveries}

We obtain two upward–closed selections: $\widehat A_{R\to B}^{\text{bin}}\subseteq \mathcal{U}_{\text{bin}}=\{0,1\}^{10}
\quad\text{and}\quad
\widehat A_{B\to R}^{\text{freq}}\subseteq \mathcal{U}_{\text{freq}}=\mathcal{L}_1\times\cdots\times \mathcal{L}_{10}$, where $\mathcal{L}_j=\{0,1\}$ for binary items, $\{0,1,2\}$ for three–level items, and $\{0,1,2,3,4\}$ for five–level items (with larger values coded as worse for the five–level items). The red$\to$blue analysis uses binary coding, while the blue$\to$red analysis uses frequency coding.

Define the deterministic coarsening map $C:\mathcal{U}_{\text{freq}}\to\mathcal{U}_{\text{bin}}$ by $C(x)_j=\mathbf{1}\{x_j\ge 1\}$ for three–/five–level items under frequency coding, $ C(x)_j=x_j$ for already–binary items. For $B\subseteq \mathcal{U}_{\text{bin}}$, define the lifting operator $\mathcal{L}(B)=\{\,x\in\mathcal{U}_{\text{freq}}:\ C(x)\in B\,\}$. For any set $S$ in either domain, let $\mathsf{up}(S)$ denote its upward closure under the coordinate–wise order in that domain.

\subsubsection{Replicable findings}
We define the \emph{replicable} set at the frequency resolution as
\[
\mathrm{Rep}\;=\;\mathsf{up}\!\Bigl(\,\widehat A_{B\to R}^{\text{freq}}\ \cap\ \mathcal{L}\!\bigl(\widehat A_{R\to B}^{\text{bin}}\bigr)\Bigr)\ \subseteq\ \mathcal{U}_{\text{freq}}.
\]
Equivalently, a frequency–coded profile $x$ is replicable iff $x\in\widehat A_{B\to R}^{\text{freq}}$ and its binary collapse $C(x)\in\widehat A_{R\to B}^{\text{bin}}$. If $x\in \mathrm{Rep}$ and $x'\succeq x$ (adding ACEs or increasing any frequency level), then $x'\in \mathrm{Rep}$. Thus we report the corners (minimal elements) of $\mathrm{Rep}$ in $\mathcal{U}_{\text{freq}}$.

For example, consider the binary corner selected in the red$\to$blue analysis:\\

$b=\{\texttt{ACESEX}{=}1,\ \text{all other ACEs}{=}0\}\in \widehat A_{R\to B}^{\text{bin}}$. Suppose the blue$\to$red analysis (under frequency coding) selects the corner $x=\{\texttt{ACESEX}{=}2,\ \text{all other ACEs}{=}0\}\in \widehat A_{B\to R}^{\text{freq}}$, where \texttt{ACESEX} has levels $\{0,1,2\}$ with larger values worse. The coarsening map $C(\cdot)$ collapses frequency to presence/absence, so $C(x)=\{\texttt{ACESEX}{=}1,\ \text{all other ACEs}{=}0\}=b$. Because $x\in \widehat A_{B\to R}^{\text{freq}}$ and $C(x)=b\in \widehat A_{R\to B}^{\text{bin}}$, we have $x\in \widehat A_{B\to R}^{\text{freq}}\ \cap\ \mathcal{L}\!\bigl(\widehat A_{R\to B}^{\text{bin}}\bigr)$, hence $x\in \mathrm{Rep}$ by definition. By monotonicity under frequency coding, any profile $x'\succeq x$ (e.g., \texttt{ACESEX}${=}2$ with \texttt{ACESWEAR}${=}1$, or \texttt{ACESEX}${=}2$ with higher levels on other ACEs) is also in $\mathrm{Rep}$ and will appear via the upward closure. Intuitively, binary coding collapses intensity. For an item with frequency levels (e.g., \texttt{ACESEX}$\in\{0,1,2\}$, larger is worse), the binary event \(\texttt{ACESEX}{=}1\) stands for the family \(\{\texttt{ACESEX}\ge 1\}\) under frequency coding. Monotonicity gives, for any fixed values of the other ACEs, $\eta(\texttt{ACESEX}{=}2,\ \text{others})\ \ge\ \eta(\texttt{ACESEX}{=}1,\ \text{others})\ \ge\ \eta(\texttt{ACESEX}{=}0,\ \text{others})$.  Hence, if a binary corner that requires \(\texttt{ACESEX}{=}1\) exceeds the threshold \(\tau\), then the refined pattern with \(\texttt{ACESEX}{=}2\) also exceeds \(\tau\). 

Below is a non–replicable contrast. Suppose instead the blue$\to$red analysis selects $y=\{\texttt{ACESEX}{=}1,\ \text{all other ACEs}{=}0\}\in \widehat A_{B\to R}^{\text{freq}}$, but the red$\to$blue analysis selects only the binary corner $b'=\{\texttt{ACESEX}{=}1,\ \texttt{ACEDIVRC}{=}1,\ \text{others}{=}0\}\in \widehat A_{R\to B}^{\text{bin}}$. Then $C(y)=\{\texttt{ACESEX}{=}1,\ \text{others}{=}0\}\\\notin \widehat A_{R\to B}^{\text{bin}}$ (because $b'$ also requires \texttt{ACEDIVRC}${=}1$). Hence $y\notin \mathcal{L}\!\bigl(\widehat A_{R\to B}^{\text{bin}}\bigr)$ and $y\notin \mathrm{Rep}$, replicability fails.

We summarize $\mathrm{Rep}$ by its corner combinations under frequency coding (see Table~\ref{tab:rep} for details). The set of replicable corners at the frequency resolution is identical to the Blue$\to$Red frequency corners. 
This follows from the two binary corners produced by the Red$\to$Blue analysis: $B_1=\{\texttt{ACEDEPRS}=1\}$,   $B_2=\{\texttt{ACESUB}=1,\ \texttt{ACEDIVRC}=1,\ \texttt{ACESWEAR}=1,\ \texttt{ACESEX}=1\}$,
with all other items equal to~0. 
Let $C(x)=\mathbf 1\{x\ge 1\}$ be the frequency$\to$binary map applied component-wise, and recall that $\mathrm{Rep}\;=\;\widehat A^{\text{freq}}_{B\to R}\ \cap\ \mathcal{L}\!\bigl(\widehat A^{\text{bin}}_{R\to B}\bigr)$, $\mathcal{L}(B)=\{x:\ C(x)\in B\}$.

\emph{Case A (\texttt{ACEDEPRS}$\ge 1$).}  
Any Blue$\to$Red frequency corner $x$ with \texttt{ACEDEPRS}$\ge 1$ satisfies $C(x)=B_1$. 
Since $B_1\in\widehat A^{\text{bin}}_{R\to B}$, we have $x\in\mathcal{L}\!\bigl(\widehat A^{\text{bin}}_{R\to B}\bigr)$ and thus $x\in\mathrm{Rep}$. 
By monotonicity, every $x'\succeq x$ (adding ACEs or raising any frequency) also lies in $\mathrm{Rep}$.

\emph{Case B (\texttt{ACEDEPRS}$=0$).}  
Each remaining Blue$\to$Red frequency corner $x$ contains 
\(\texttt{ACESUB}\ge 1,\ \texttt{ACEDIVRC}\ge 1,\ \texttt{ACESWEAR}\ge 1,\ \texttt{ACESEX}\ge 1\).
Hence $C(x)\succeq B_2$ and in particular $C(x)\in\widehat A^{\text{bin}}_{R\to B}$. 
Therefore $x\in\mathcal{L}\!\bigl(\widehat A^{\text{bin}}_{R\to B}\bigr)$ and $x\in\mathrm{Rep}$.

Because these two corners cover all Blue$\to$Red frequency corners, the replicable set coincides with the Blue$\to$Red frequency selection. Therefore, the replicable higher-risk subgroup covers $6{,}166$ profiles, i.e., $19.03\%$ of the trequency grid, and in the BRFSS 2023 sample with frequency coding ($N=49{,}547$), the rule flags $4{,}364$ respondents, i.e., $8.81\%$ of individuals.

\begin{landscape}
\begin{table}
\small\sf\centering
\small
\caption{Corner Combinations of the \emph{Replicable} High-risk Subgroup under \emph{Frequency} Coding}
\begin{tabular}{@{}c *{10}{c}@{}}
\toprule
 & ACEDEPRS & ACESUB & ACEPRISN & ACEDIVRC & ACEPUNCH & ACEHURT1 & ACESWEAR & ACESEX & ACEADSAF & ACEADNED \\ 
\midrule\addlinespace[2.5pt]
Corner 1 & 1 & 0 & 0 & 0 & 0 & 0 & 2 & 1 & 0 & 0 \\ 
Corner 2 & 1 & 1 & 1 & 1 & 0 & 0 & 0 & 2 & 0 & 0 \\ 
Corner 3 & 1 & 1 & 0 & 1 & 0 & 1 & 1 & 2 & 0 & 0 \\ 
Corner 4 & 1 & 1 & 0 & 0 & 0 & 0 & 2 & 0 & 1 & 0 \\ 
Corner 5 & 1 & 1 & 1 & 0 & 0 & 0 & 0 & 2 & 1 & 0 \\ 
Corner 6 & 1 & 1 & 0 & 1 & 0 & 0 & 0 & 2 & 1 & 0 \\ 
Corner 7 & 1 & 0 & 0 & 1 & 1 & 1 & 2 & 0 & 2 & 0 \\ 
Corner 8 & 1 & 0 & 1 & 0 & 1 & 2 & 2 & 0 & 2 & 0 \\ 
Corner 9 & 1 & 1 & 1 & 0 & 1 & 1 & 2 & 0 & 0 & 1 \\ 
Corner 10 & 1 & 1 & 0 & 1 & 0 & 1 & 0 & 2 & 0 & 1 \\ 
Corner 11 & 1 & 0 & 1 & 0 & 0 & 1 & 2 & 0 & 1 & 1 \\ 
Corner 12 & 1 & 0 & 0 & 1 & 0 & 1 & 2 & 0 & 2 & 1 \\ 
Corner 13 & 1 & 0 & 0 & 0 & 0 & 2 & 2 & 0 & 2 & 1 \\ 
Corner 14 & 0 & 1 & 0 & 1 & 1 & 1 & 2 & 2 & 1 & 0 \\ 
Corner 15 & 0 & 1 & 0 & 1 & 1 & 0 & 2 & 2 & 2 & 0 \\ 
Corner 16 & 0 & 1 & 0 & 1 & 1 & 0 & 2 & 2 & 1 & 1 \\ 
Corner 17 & 0 & 1 & 0 & 1 & 0 & 2 & 2 & 2 & 2 & 1 \\  
\bottomrule
\label{tab:rep}
\end{tabular}
\end{table}
\end{landscape}

\subsubsection{Global Discoveries}

Let $C:\mathcal{U}_{\text{freq}}\to\mathcal{U}_{\text{bin}}$ be the coarsening map and, for $B\subseteq\mathcal{U}_{\text{bin}}$, the lifting operator $\mathcal{L}(B)=\{\,x\in\mathcal{U}_{\text{freq}}:\ C(x)\in B\,\}$.

We report the global discoveries under frequency coding as
\[
G \;=\; \mathsf{up}\!\Bigl(\,\widehat A_{B\to R}^{\text{freq}} \ \cup\ \mathcal{L}\!\bigl(\widehat A_{R\to B}^{\text{bin}}\bigr)\Bigr)\ \subseteq\ \mathcal{U}_{\text{freq}}.
\]
A profile $x$ is in the global set iff it is high–risk in at least one part (blue$\to$red at the native frequency resolution, or red$\to$blue after lifting its binary discoveries). 

For example, let $b=(\texttt{ACESEX}=1,\ \text{others}=0)\in\mathcal{U}_{\text{bin}}$ be a binary corner selected in the red$\to$blue analysis, and let $x^{(2)}=(\texttt{ACESEX}=2,\ \text{others}=0)\in\mathcal{U}_{\text{freq}}$ be a frequency corner selected in the blue$\to$red analysis. The lift of $\{b\}$ is $\mathcal{L}\!\bigl(\{b\}\bigr)\;=\;\bigl\{\,x\in\mathcal{U}_{\text{freq}}:\ C(x)=b\,\bigr\}
\;=\;\bigl\{\,x:\ x_{\texttt{ACESEX}}\ge 1,\ \text{others}=0\,\bigr\}$.

Hence the global set reported under frequency coding $G\;=\;\mathsf{up}\!\Bigl(\,\widehat A_{B\to R}^{\text{freq}}\ \cup\ \mathcal{L}\!\bigl(\{b\}\bigr)\Bigr)$
contains all profiles with \(\texttt{ACESEX}\ge 1\) and other items at 0, as well as any blue–side frequency selections. By monotonicity, the frequency profile $x^{(1)}=(\texttt{ACESEX}=1,\ \text{others}=0)$ is a corner of \(G\), while \(x^{(2)}\succeq x^{(1)}\) lies above it and therefore is not a corner though it remains in \(G\) by upward closure.

We present $G$ by its corner points in $\mathcal{U}_{\text{freq}}$. Specifically, there are two corners under frequency coding: $\bigl\{
\{\texttt{ACEDEPRS}=1, \texttt{others}=0\},
\;\{\texttt{ACESUB}=1,\ \texttt{ACEDIVRC}=1,\ \texttt{ACESWEAR}=1,\ \texttt{ACESEX}=1,
\texttt{others}=0\}
\bigr\}$. On the frequency grid, $G$ covers $18{,}000$ profiles ($55.56\%$).
In the BRFSS 2023 sample with frequency coding ($N=49{,}547$), the global rule flags $9{,}842$ respondents ($19.86\%$).

\section{Simulations}
\label{simulation}

\subsection{Simulation Settings}

We generate a latent frequency–coded profile $\tilde X=(x_1,\ldots,x_{10})$ on a mixed grid:
$x_{1:4}\in\{0,1\}$ (binary), $x_{5:8}\in\{0,1,2\}$ (three–level), and $x_{9:10}\in\{0,1,2,3,4\}$ (five–level).
Coordinates are independent with fixed marginals matching the application.
We then coarsen to a binary vector by $X^{\text{bin}} \;=\; C(\tilde X), C(x)_j=\mathbf{1}\{x_j\ge 1\}$.

The outcome model is defined on the latent frequency space $\Pr(Y{=}1\mid \tilde X)=\operatorname{logit}^{-1}\!\bigl(b_0 + s\,\eta_0(\tilde X)\bigr), b_0=\operatorname{logit}(0.10)$\footnote{$\operatorname{logit}(p) \;=\; \log\!\left(\frac{p}{1-p}\right), \operatorname{logit}^{-1}(z) \;=\; \frac{1}{1+e^{-z}}\, .$}.
The signal $\eta_0(\cdot)$ is monotone in each coordinate level. We consider two signal shapes for $\eta_0(\cdot)$.(i) a logistic model only with main effects (coordinate-wise monotonicity and no interactions), $\operatorname{logit}\, p(x) = \beta_0 + \sum_{j=1}^{10} \beta_j x_j,  \beta_j \ge 0 \ \text{for all } j$, so the risk is non-decreasing in each coordinate. (ii) the additive form augmented with a positive interaction term on the logit scale, $\operatorname{logit} \, p(x) = \beta_0 + \sum_{j=1}^{10} \beta_j x_j \;+\; \gamma\, x_9 x_{10}, \gamma>0$.
We choose the scale $s$ by calibration so that the superlevel mass $\Pr\bigl(\eta(\tilde X)\ge\tau\bigr)$ equals a target in $\{0.50,0.60,0.70\}$.

After simulating $(\tilde X,Y)$ we record both codings $(\tilde X,\; X^{\text{bin}})$ and split observations into blue ($45\%$) and red ($55\%$). We fix the ISS threshold at $\tau=0.20$. The scale $s$ is calibrated by root–finding on the latent
frequency distribution so that the superlevel mass satisfies $\Pr\!\bigl(\eta(\tilde X)\ge \tau\bigr)\in\{0.50,0.60,0.70\}$.
Sample sizes are $n\in\{10{,}000,20{,}000,30{,}000,40{,}000,50{,}000,60{,}000\}$. The familywise error rate is controlled at $\alpha=0.05$. For each design cell we run $R=100$ replications with fixed seeds.

\subsubsection{Part 1: Parent Selection Rules}We first run the simulation on binary coded data to compare different parent selection rules.
\begin{enumerate}\itemsep2pt
\item \textbf{Nearest–cover baseline.} Attach each node to a strictly dominating cover that minimizes the $\ell_\infty$ distance on $\{0,1\}^{10}$; break ties at random (matches the default in the ISS R package).
\item \textbf{Evidence–guided rule.} Among screened covers, attach to the cover whose \emph{screening} $p$–value (computed in the opposite part) is smallest; break ties at random.
\end{enumerate}
This corresponds to our design in the pre-specified plan. Both ways can be fully automated. In other words, we don't need to explore the data to organize the structure of the polyforest. Therefore, in this module of simulation, we don't include EDA. We run the pre–specified cross–screening in both directions on the binary grid, hold $\tau$ and the screening cutoff fixed, and vary only how we attach at most one parent (cover) per node in the validation part. For reporting, we first summarize with: (i) \emph{Intersection (replicable)}: $\widehat A_{\cap} \;=\; \widehat A_{\text{B}\to\text{R}}^{\text{bin}} \cap \widehat A_{\text{R}\to\text{B}}^{\text{bin}}$. Profiles in $\widehat A_{\cap}$ are rejected in both parts, hence are replicable. (ii) \emph{Union (global discovery)}: $\widehat A_{\cup} \;=\; \widehat A_{\text{B}\to\text{R}}^{\text{bin}} \cup \widehat A_{\text{R}\to\text{B}}^{\text{bin}}$. Profiles in $\widehat A_{\cup}$ are rejected in at least one part, so the global null is rejected.

For each profile $x\in\mathcal{U}_{\text{freq}}$, we compute
$\Pr(Y{=}1\mid \tilde X{=}x)$ under the model and mark $x$ as truly high–risk if this
probability is at least $\tau$.
The frequency–level oracle set is
$\mathcal{T}_{\text{freq}}=\mathsf{up}\{\,x:\Pr(Y{=}1\mid \tilde X{=}x)\ge\tau\,\}$.
We compute the population mass $\pi_{\text{freq}}(x)$ of each profile from the
product marginals used by the DGP. For the binary coded datasets,let $C:\mathcal{U}_{\text{freq}}\!\to\!\mathcal{U}_{\text{bin}}$ be the deterministic
coarsening with $C(x)_j=\mathbf{1}\{x_j\ge 1\}$ for nonbinary items and
$C(x)_j=x_j$ for already–binary items.
We define the binary oracle set by coarsening the frequency truth and taking
upward closure:
$\mathcal{T}_{\text{bin}}=\mathsf{up}\bigl(C(\mathcal{T}_{\text{freq}})\bigr)$.
The mass of a binary profile $b$ is the induced mass
$\pi_{\text{bin}}(b)=\sum_{x:\,C(x)=b}\pi_{\text{freq}}(x)$. These quantities are used only for evaluation.
We summarize performance by \emph{average regret}-
the total population mass of truly high–risk profiles not covered by the selected
set, divided by the total grid mass.
We also report empirical FWER, defined as the fraction of simulation runs with at
least one false inclusion (i.e., a profile with $\Pr(Y{=}1\mid \tilde X{=}x)<\tau$ entering $\hat A$).

\subsubsection{Part 2: Coding Choices}Secondly, inspired by the change of coding criteria during our EDA on the red part, we set up the simulation to compare three different ways of coding.
\begin{enumerate}\itemsep2pt
\item \textbf{ISS on ACE score.}
Let $Z(x)=\sum_{j=1}^{10}\mathbf{1}\{x_j\ge 1\}$ be the ACE score.
The domain is the totally ordered set $\{0,1,\dots,10\}$ with $0\prec1\prec\cdots\prec10$.
For each level $z$, test $H_{0}(z):\ \eta(Z{=}z)<\tau
\quad\text{vs.}\quad
H_{1}(z):\ \eta(Z{=}z)\ge\tau$, under the monotonicity constraint $\eta(z)\le \eta(z')$ for $z\le z'$.
ISS on this chain yields a data–driven threshold $\widehat z_0$ (the smallest rejected level).
The selected set at the profile level is $\{x:\ Z(x)\ge \widehat z_0\}$.

\item \textbf{ISS with binary coding.}
Collapse each item to presence/absence $\mathbf{1}\{x_j\ge 1\}$ and run ISS on the $d{=}10$–dimensional product order, as specified earlier.

\item \textbf{ISS with frequency coding.}
Use the original ordinal levels (three–/five–level where applicable), and run ISS on the full grid.
\end{enumerate}

For each observation $i=1,\dots,n$ with profile $X_i$, define the oracle label $T_i \;=\; \mathbf{1}\!\left\{\, \Pr(Y{=}1 \mid X_i)\ \ge\ \tau \,\right\}$. Given a method $m$ that selects a set of profiles $\widehat A^{(m)}$, define the predicted label $\widehat T_i^{(m)} \;=\; \mathbf{1}\!\left\{\, X_i \in \widehat A^{(m)} \,\right\}$. Evaluate:
\begin{itemize}\itemsep2pt
\item \emph{ISS on frequency coding}: check $X_i \in \widehat A^{(\mathrm{freq})}$ directly.
\item \emph{ISS on binary coding}: collapse $X_i$ to $C(X_i)_j=\mathbf{1}\{X_{ij}\ge1\}$ and check $C(X_i)\in \widehat A^{(\mathrm{bin})}$.
\item \emph{ISS on ACE score}: compute $Z(X_i)=\sum_j \mathbf{1}\{X_{ij}\ge1\}$ and check $Z(X_i)\ge \widehat z_0$.
\end{itemize}

From $\{T_i,\widehat T_i^{(m)}\}_{i=1}^n$ we compute $\mathrm{TP}=\sum \mathbf{1}\{T_i=1,\widehat T_i^{(m)}=1\}$, $\mathrm{FP}=\sum \mathbf{1}\{T_i=0,\widehat T_i^{(m)}=1\}$, $\mathrm{FN}=\sum \mathbf{1}\{T_i=1,\widehat T_i^{(m)}=0\}$, $\mathrm{TN}=\sum \mathbf{1}\{T_i=0,\widehat T_i^{(m)}=0\}$, and report $\mathrm{Sensitivity}=\frac{\mathrm{TP}}{\mathrm{TP}+\mathrm{FN}}$, $\mathrm{Specificity}=\frac{\mathrm{TN}}{\mathrm{TN}+\mathrm{FP}}$,
$\text{positive predictive value}(\mathrm{PPV})=\frac{\mathrm{TP}}{\mathrm{TP}+\mathrm{FP}}$ and negative predictive value($\mathrm{NPV}$)= $\frac{\mathrm{TN}}{\mathrm{TN}+\mathrm{FN}}$.

\subsection{Results}

\begin{figure}
\centering
\includegraphics[width=1\linewidth]{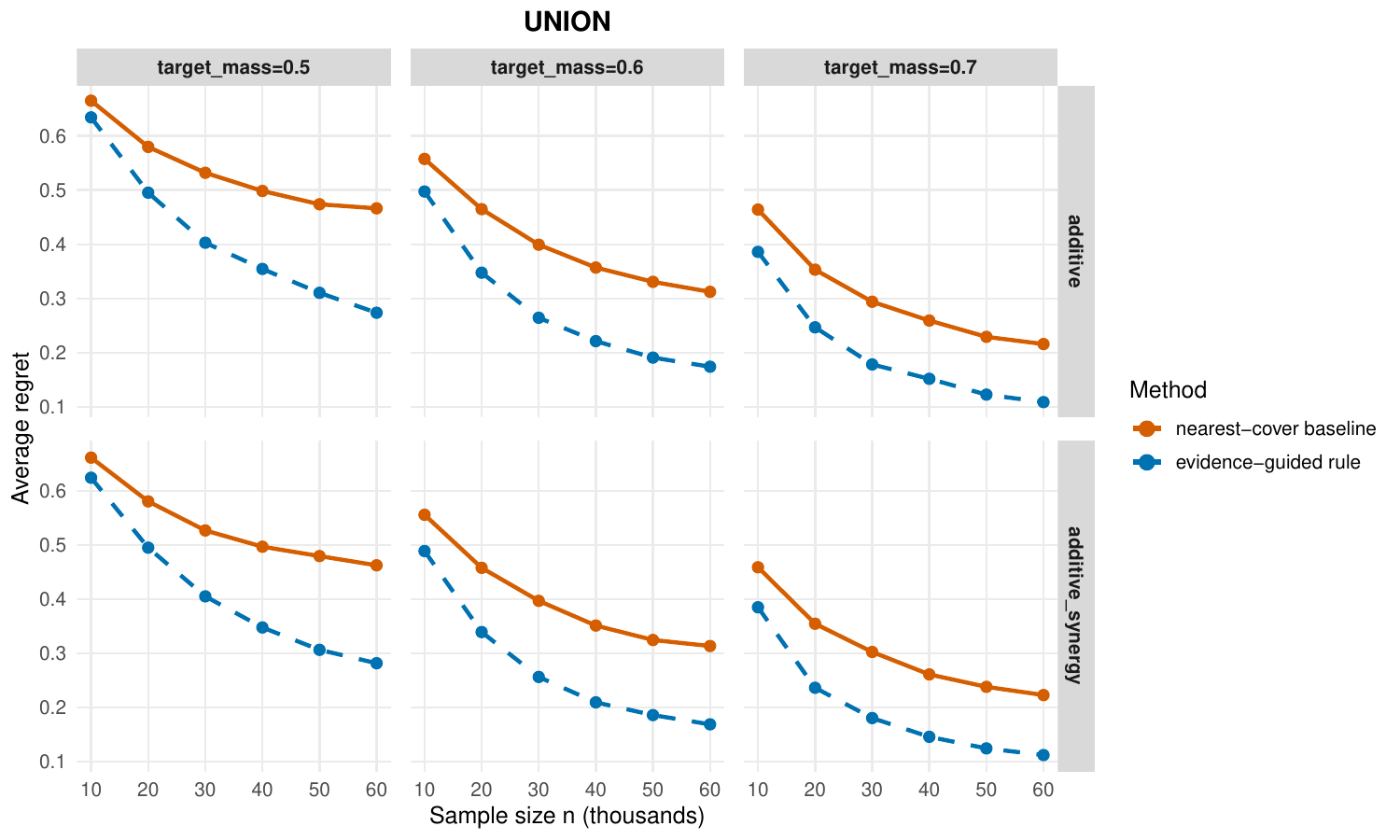}
\caption{\textbf{Union (global discovery).} Average regret versus sample size $n$ for the union of higher-risk groups identified in the two parts. Panels vary the target mass and the functional form $\eta(\cdot)$. Solid orange: nearest–cover baseline, parent chosen by minimum $\ell_\infty$ distance (the default in the \texttt{ISS} R package). Dashed blue: evidence–guided parent selection, parent chosen as the screened cover with the smallest \emph{opposite–part} screening $p$–value). Lower is better.}
\label{fig:union}
\end{figure}

\begin{figure}
\centering
\includegraphics[width=1\linewidth]{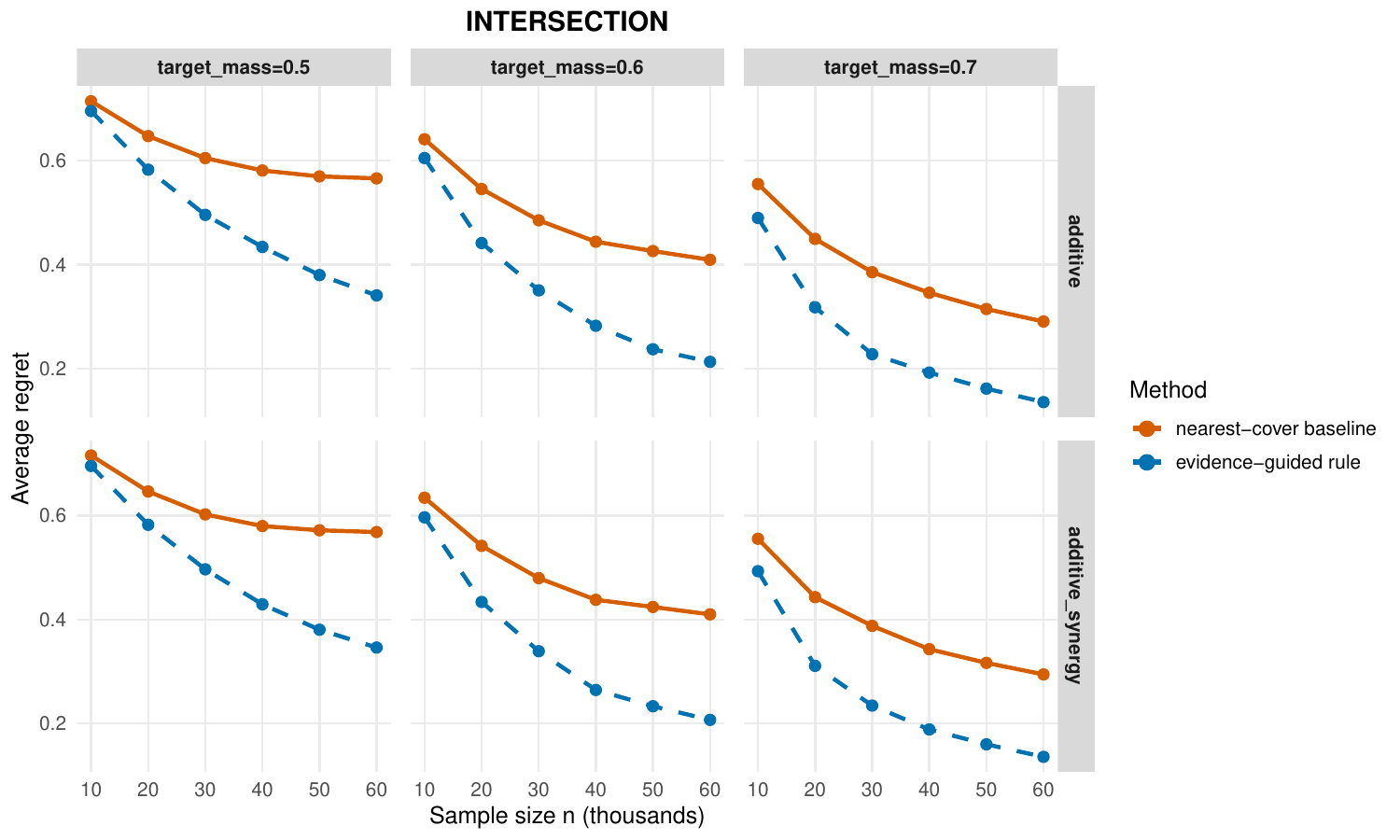}
\caption{\textbf{Intersection (replicable discovery).} Average regret versus sample size $n$ for the intersection of higher-risk groups identified in the two parts. Layout and line styles match Figure~\ref{fig:union}.}
\label{fig:intersection}
\end{figure}

Figures~\ref{fig:union} and \ref{fig:intersection} summarize performance across sample sizes, target masses, and data-generating mechanisms. The evidence-guided polyforest achieves uniformly lower average regret than the nearest-cover baseline. Importantly, the ordering of methods is unchanged in both the union (global discovery of higher-risk group) and the intersection (replicable higher-risk group), and for all different settings of data generating process (DGP). As expected, empirical FWER remains below the controlled level $\alpha=0.05$ across all configurations. Thus, the evidence-guided construction of polyforest improves coverage of truly high-risk combinations without sacrificing error control.

\begin{figure}
\centering
\includegraphics[width=1\linewidth]{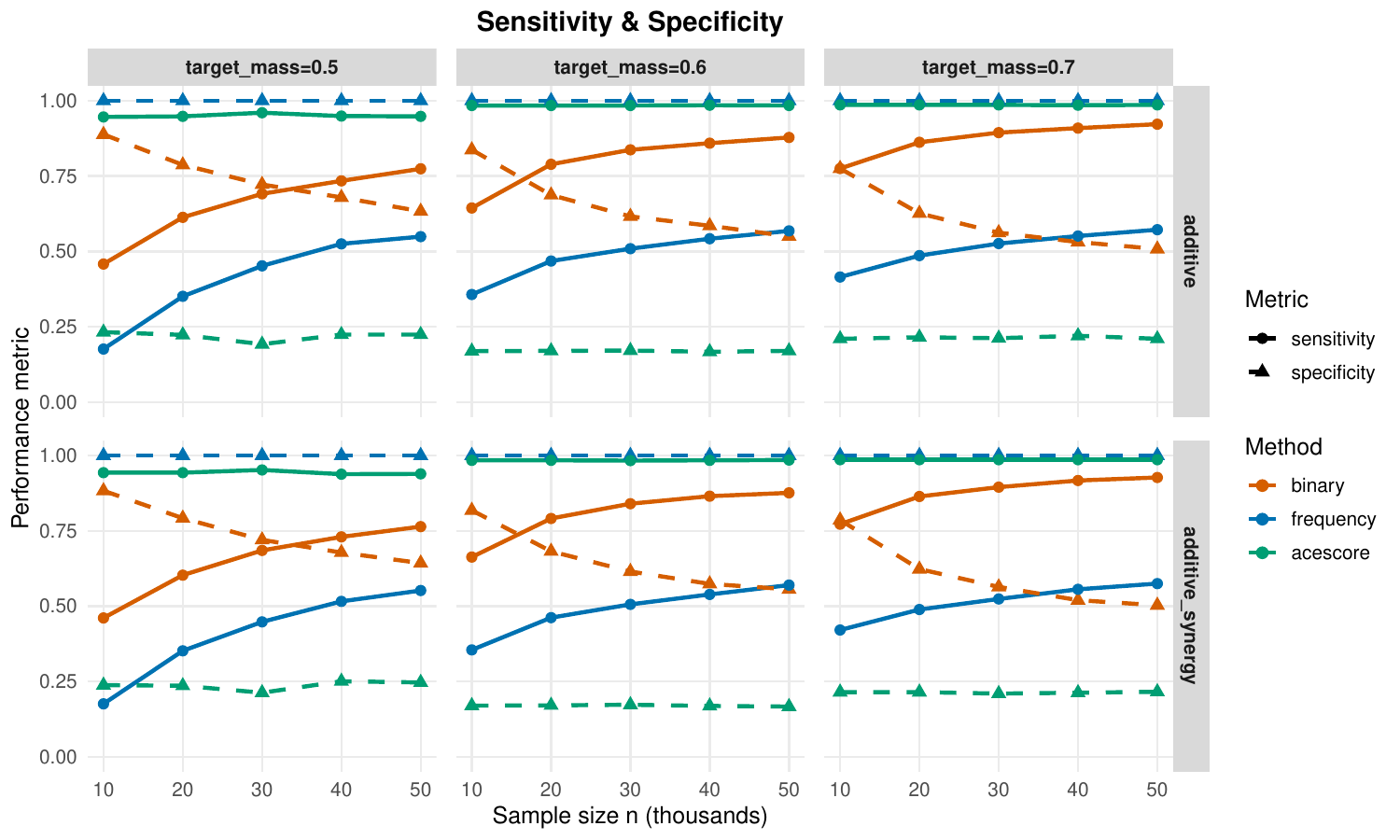}
\caption{Sensitivity and specificity by coding choice. Curves plot each metric against sample size. Columns vary the target superlevel mass; rows vary the DGP. Three implementations of ISS are shown: ACE–score, binary coding, and frequency coding.}
\label{fig:sensspec}
\end{figure}

\begin{figure}
\centering
\includegraphics[width=1\linewidth]{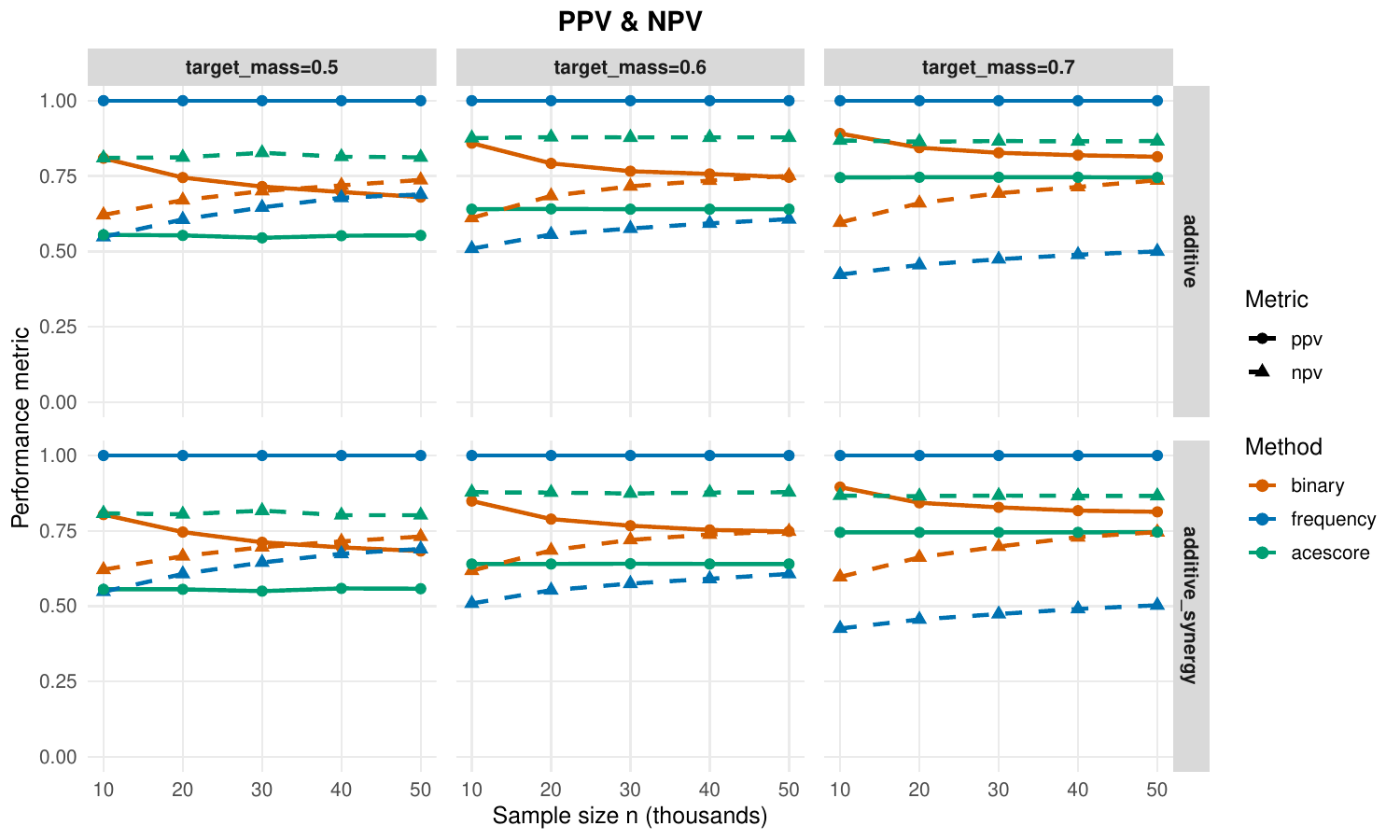}
\caption{Positive predictive value (PPV) and negative predictive value (NPV) by coding choice. Layout and line styles/colors match Figure~\ref{fig:sensspec}.}
\label{fig:ppvnpv}
\end{figure}

Figures~\ref{fig:sensspec}–\ref{fig:ppvnpv} show a clear, stable pattern across data‐generating mechanisms and target masses.
First, specificity and PPV are consistently highest under frequency-coded ISS, with only mild variation as $n$ grows.
Both binary and ACE score implementations remain well below the frequency curves on these two metrics throughout, reflecting structural information loss. Keeping frequency levels retains both which ACE items are present and how often, allowing ISS to rule out low–intensity look–alikes and thereby reduce false positives while collapsing to binary items or to a single score discards exactly this information, which limits attainable specificity and PPV. This is intrinsic to the coding. 

Second, sensitivity and NPV exhibit the expected trade–off respectively with specificity and PPV.
The ACE score version attains the largest sensitivity/NPV at small and moderate $n$ because it triggers on many coarse profiles, whereas binary is intermediate and frequency starts lower but improves steadily with sample size.
Indeed, as $n$ increases, sensitivity for both binary and frequency coding rises (greater power), while specificity and PPV remain comparatively stable for all three methods.

From a screening standpoint, our aim is to identify a replicable higher–risk subgroup that can be used as a practical trigger for follow-up assessment or intervention under finite program capacity. Prior work (e.g., \cite{meehan2022poor}) cautions that implementing ACE screening via ACE score thresholds can have detrimental consequences for resource allocation because low specificity admits many false positives. In our setting, using binary collapse or ACE score similarly enrolls many low-intensity or off-pattern profiles, which manifests as lower specificity. By contrast, at the same prespecified risk threshold $\tau$, frequency-coded ISS consistently yields higher specificity and PPV, while its sensitivity improves as sample size increases. The resulting rule is also operationally clear that it specifies which items and the minimum levels required. Taken together with the blue–part EDA, our simulations quantify and corroborate the advantage of adopting frequency coding.

\section{External Comparison}
\label{comparison}

We benchmark simple ACE–score cutoffs against the replicable higher-risk combinations we just identified on an external held-out set - BRFSS 2022. We keep the outcome definition consistent with earlier sections. Let $Y{=}1$ if the respondent reports ever being told they had a depressive disorder, and $0$ otherwise (variable \texttt{ADDEPEV3} in BRFSS). We reuse the same ACE item construction and the same frequency coding as in the application.

Specifically, we compare two screening rules.
\begin{enumerate}\itemsep3pt
\item \textbf{ACE–score threshold.} Let $Z(x)=\sum_{j=1}^{10}\mathbf{1}\{x_j\ge1\}$ be the ACE count under frequency coding. For $K\in\{1, 2,3,4,5,6,7,8,9\}$, classify $x$ as high–risk iff $Z(x)\ge K$.
\item \textbf{Replicable higher-risk subgroup} Let $\mathrm{Rep}\subseteq\mathcal{U}_{\text{freq}}$ be the frequency–coded replicable set defined in the application. Classify $x$ as high–risk iff $x\in\mathrm{Rep}$.
\end{enumerate}

We report standard sample–level metrics. Let $R(X)\in\{0,1\}$ be a rule’s classification on an individual profile $X$: $\text{PPR}=\mathbb{E}[R(X)]$, $\text{Sensitivity}=\mathbb{E}[R(X)\mid Y{=}1]$, $\text{Specificity}=\mathbb{E}[1{-}R(X)\mid Y{=}0]$, $\text{PPV}=\mathbb{P}(Y{=}1\mid R(X){=}1)$, $\text{NPV}=\mathbb{P}(Y{=}0\mid R(X){=}0)$. We estimate all quantities by sample proportions on BRFSS 2022. 

We construct the BRFSS 2022 analysis set using the same inclusion rules as in the application. In total, the analysis set includes $27,652$ samples. We evaluate each $K\in\{1,2,3,4,5,6,7,8,9\}$ and the replicable higher-risk combinations on the identical set of respondents. 

Table~\ref{tab:screen2022} compares ACE score cutoffs with our replicable rule on BRFSS~2022. Motivated by the Neyman--Pearson paradigm (hold the Type I error fixed and maximize power), we align rules at the same specificity and then compare sensitivity. 
At specificity $=0.95$ (false–positive rate $=0.05$), the ACE score cutoff $\ge 7$ attains sensitivity $=0.19$, PPV $=0.49$, NPV $=0.83$, and PPR $=0.08$. 
Our replicable subgroup achieves the same specificity $=0.95$ with higher sensitivity $=0.24$ (a $26\%$ relative gain),  and also higher PPV $=0.53$ and slightly higher NPV $=0.84$. 
Lower ACE cutoffs increase sensitivity but sharply reduce specificity and PPV, while higher cutoffs increase PPV at the cost of very low sensitivity and very small PPR. Under the fixed--specificity comparison above, our replicable higher‑risk subgroup performs better because it catches more truly at‑risk individuals. In other words, using our replicable higher-risk subgroup controls false positives to the certain extent so resources aren’t wasted and simultaneously misses fewer true cases by using specific item and frequency information rather than a cumulative score or binary items. Besides, in BRFSS 2022, matching the subgroup’s specificity requires an ACE score cutoff of $\ge 7$. However, the commonly used $\ge 4$ cutoff is much less specific and would admit many more low-risk respondents, which may have detrimental consequences for resource allocation.

\begin{table}
\centering\small
\caption{Screening metrics on BRFSS 2022 by rule.}
\label{tab:screen2022}
\begin{tabular}{lccccc}
\toprule
Rule & PPR & Sensitivity & Specificity & PPV & NPV \\
\midrule
ACE score $\ge 1$ & 0.67 & 0.84 & 0.38 & 0.25 & 0.91 \\
ACE score $\ge 2$ & 0.45 & 0.68 & 0.61 & 0.30 & 0.89 \\
ACE score $\ge 3$ & 0.31 & 0.55 & 0.74 & 0.34 & 0.87 \\
ACE score $\ge 4$ & 0.23 & 0.45 & 0.83 & 0.39 & 0.86 \\
ACE score $\ge 5$ & 0.17 & 0.35 & 0.88 & 0.41 & 0.85 \\
ACE score $\ge 6$ & 0.12 & 0.27 & 0.92 & 0.45 & 0.84 \\
\textbf{ACE score $\ge 7$} & \textbf{0.08} & \textbf{0.19} & \textbf{0.95} & \textbf{0.49} & \textbf{0.83} \\
ACE score $\ge 8$ & 0.05 & 0.12 & 0.97 & 0.52 & 0.82 \\
ACE score $\ge 9$ & 0.02 & 0.06 & 0.99 & 0.57 & 0.81 \\
\textbf{Replicable higher-risk subgroup} & \textbf{0.09} & \textbf{0.24} & \textbf{0.95} & \textbf{0.53} & \textbf{0.84} \\
\bottomrule
\end{tabular}
\end{table}

\section{Conclusions}
\label{conclusion}

We used data turnover combined with isotonic subgroup selection (ISS) to identify a replicable higher–risk subgroup of ACE exposure patterns. ISS’s monotonicity assumption matches the question of how ACEs relate to adult depression: if a profile $x$ is made more adverse to $x'$ by adding ACE items or increasing their frequency ($x\preceq x'$), the depression risk should not decrease, i.e., $\eta(x)\le \eta(x')$. This implies that the selected subgroup is upward–closed and can be summarized by its minimal corner combinations (i.e., the weakest combinations that still exceed the threshold $\tau$) making the result transparent and easy to implement in screening. The data turnover framework allows us to do exploration, confirmation, and replication with only one team of researchers. On the exploratory side we examined coding choices and found that incorporating frequency levels for selected ACE items is well–matched to the ACE context. On the confirmatory/replication side, cross–screening and validation preserved strong error control and improved effective power of ISS by focusing testing on promising candidates and organizing the structure of the polyforest to avoid bottlenecks. 

Exploration contributed substantively in our work. Frequency coding preserves which ACE items appear and how often, revealing boundary combinations that binary collapse cannot. This produced concrete, pattern–based triggers that replicated across splits. Replicability, in turn, increases credibility for downstream use for screening and supports transport to external data.

We then evaluated this replicable higher-risk group as screening rules on an external held–out set (BRFSS 2022). Under a fixed–specificity comparison (matching specificity at 0.95), the replicable subgroup achieved higher sensitivity (increase by around $26\%$) and higher PPV than the ACE score threshold needed to reach the same specificity (here, ACE score $\ge 7$). This shows that one can control false positives to the same extent yet miss fewer truly at–risk individuals by using specific item and frequency patterns rather than a cumulative score or binary items.

Our design prespecifies a risk threshold $\tau$ that defines higher risk. In this study we set $\tau$ to approximately double the odds of depression relative to the no ACE baseline. The same framework permits other choices of $\tau$ to match program goals and capacity: a lower $\tau$ yields broader screening (higher sensitivity and lower specificity), whereas a higher $\tau$ yields narrower screening (lower sensitivity and higher specificity).

In BRFSS 2022 our replicable subgroup attains sensitivity $=0.24$ at specificity $=0.95$. For depression screening this is a favorable operating point relative to ACE score cutoffs, but it also underscores that ACEs are only one component of risk. Stronger screening pipelines will likely need to integrate additional predictors (e.g., sociodemographics, current symptoms, stressors, social determinants) alongside ACE patterns. In addition, heterogeneity likely matters for the effect of ACEs on depression. Selected patterns may vary by sex, race/ethnicity, and age. Extending to assess subgroup–specific rules with multiplicity control and to monitor stability across years and regions is a natural direction for future work.

\section*{Acknowledgments}
We thank Prof. Jacob Bien for the insightful conversations and guidance during the early stages of this project, these discussions greatly helped shape the direction and development of our work.

\newpage

% \bibliography{../references.bib}
\bibliography{references.bib}

@article{felitti1998relationship,
  title={Relationship of childhood abuse and household dysfunction to many of the leading causes of death in adults: The Adverse Childhood Experiences (ACE) Study},
  author={Felitti, Vincent J and Anda, Robert F and Nordenberg, Dale and Williamson, David F and Spitz, Alison M and Edwards, Valerie and Marks, James S},
  journal={American journal of preventive medicine},
  volume={14},
  number={4},
  pages={245--258},
  year={1998},
  publisher={Elsevier}
}

@article{dube2003impact,
  title={The impact of adverse childhood experiences on health problems: evidence from four birth cohorts dating back to 1900},
  author={Dube, Shanta R and Felitti, Vincent J and Dong, Maxia and Giles, Wayne H and Anda, Robert F},
  journal={Preventive medicine},
  volume={37},
  number={3},
  pages={268--277},
  year={2003},
  publisher={Elsevier}
}

@article{anda2006enduring,
  title={The enduring effects of abuse and related adverse experiences in childhood: A convergence of evidence from neurobiology and epidemiology},
  author={Anda, Robert F and Felitti, Vincent J and Bremner, J Douglas and Walker, John D and Whitfield, CH and Perry, Bruce D and Dube, Sh R and Giles, Wayne H},
  journal={European archives of psychiatry and clinical neuroscience},
  volume={256},
  pages={174--186},
  year={2006},
  publisher={Springer}
}

@article{chapman2004adverse,
  title={Adverse childhood experiences and the risk of depressive disorders in adulthood},
  author={Chapman, Daniel P and Whitfield, Charles L and Felitti, Vincent J and Dube, Shanta R and Edwards, Valerie J and Anda, Robert F},
  journal={Journal of affective disorders},
  volume={82},
  number={2},
  pages={217--225},
  year={2004},
  publisher={Elsevier}
}

@article{anda2002adverse,
  title={Adverse childhood experiences, alcoholic parents, and later risk of alcoholism and depression},
  author={Anda, Robert F and Whitfield, Charles L and Felitti, Vincent J and Chapman, Daniel and Edwards, Valerie J and Dube, Shanta R and Williamson, David F},
  journal={Psychiatric services},
  volume={53},
  number={8},
  pages={1001--1009},
  year={2002},
  publisher={Am Psychiatric Assoc}
}

@article{williamson2002body,
  title={Body weight and obesity in adults and self-reported abuse in childhood},
  author={Williamson, David F and Thompson, Theodore J and Anda, Robert F and Dietz, William H and Felitti, Vincent},
  journal={International journal of obesity},
  volume={26},
  number={8},
  pages={1075--1082},
  year={2002},
  publisher={Nature Publishing Group}
}

@article{dong2004interrelatedness,
  title={The interrelatedness of multiple forms of childhood abuse, neglect, and household dysfunction},
  author={Dong, Maxia and Anda, Robert F and Felitti, Vincent J and Dube, Shanta R and Williamson, David F and Thompson, Theodore J and Loo, Clifton M and Giles, Wayne H},
  journal={Child abuse \& neglect},
  volume={28},
  number={7},
  pages={771--784},
  year={2004},
  publisher={Elsevier}
}

@article{felitti2002relationship,
  title={The relationship of adverse childhood experiences to adult health: Turning gold into lead/Belastungen in der Kindheit und Gesundheit im Erwachsenenalter: die Verwandlung von Gold in Blei},
  author={Felitti, Vincent J},
  journal={Zeitschrift f{\"u}r Psychosomatische Medizin und Psychotherapie},
  volume={48},
  number={4},
  pages={359--369},
  year={2002},
  publisher={Vandenhoeck \& Ruprecht GmbH \& Co. KG G{\"o}ttingen}
}

@article{bhan2014childhood,
  title={Childhood adversity and asthma prevalence: evidence from 10 US states (2009--2011)},
  author={Bhan, Nandita and Glymour, M Maria and Kawachi, Ichiro and Subramanian, SV4212798},
  journal={BMJ Open Respiratory Research},
  volume={1},
  number={1},
  pages={e000016},
  year={2014},
  publisher={Archives of Disease in childhood}
}

@article{waehrer2020disease,
  title={Disease burden of adverse childhood experiences across 14 states},
  author={Waehrer, Geetha M and Miller, Ted R and Silverio Marques, Sara C and Oh, Debora L and Burke Harris, Nadine},
  journal={PLoS one},
  volume={15},
  number={1},
  pages={e0226134},
  year={2020},
  publisher={Public Library of Science San Francisco, CA USA}
}

@article{hughes2017effect,
  title={The effect of multiple adverse childhood experiences on health: a systematic review and meta-analysis},
  author={Hughes, Karen and Bellis, Mark A and Hardcastle, Katherine A and Sethi, Dinesh and Butchart, Alexander and Mikton, Christopher and Jones, Lisa and Dunne, Michael P},
  journal={The Lancet public health},
  volume={2},
  number={8},
  pages={e356--e366},
  year={2017},
  publisher={Elsevier}
}

@article{merrick2019vital,
  title={Vital signs: estimated proportion of adult health problems attributable to adverse childhood experiences and implications for prevention—25 states, 2015--2017},
  author={Merrick, Melissa T},
  journal={MMWR. Morbidity and mortality weekly report},
  volume={68},
  year={2019}
}

@article{gilbert2015childhood,
  title={Childhood adversity and adult chronic disease: an update from ten states and the District of Columbia, 2010},
  author={Gilbert, Leah K and Breiding, Matthew J and Merrick, Melissa T and Thompson, William W and Ford, Derek C and Dhingra, Satvinder S and Parks, Sharyn E},
  journal={American journal of preventive medicine},
  volume={48},
  number={3},
  pages={345--349},
  year={2015},
  publisher={Elsevier}
}

@article{bekerman2024planning,
  title={Planning for gold: Hypothesis screening with split samples for valid powerful testing in matched observational studies},
  author={Bekerman, William and Dalal, Abhinandan and Del Ninno, Carlo and Small, Dylan S},
  journal={Biometrika},
  pages={asaf078},
  year={2025},
  publisher={Oxford University Press}
}

@article{cox1975note,
  title={A note on data-splitting for the evaluation of significance levels},
  author={Cox, David R},
  journal={Biometrika},
  pages={441--444},
  year={1975},
  publisher={JSTOR}
}

@article{heller2009split,
  title={Split samples and design sensitivity in observational studies},
  author={Heller, Ruth and Rosenbaum, Paul R and Small, Dylan S},
  journal={Journal of the American Statistical Association},
  volume={104},
  number={487},
  pages={1090--1101},
  year={2009},
  publisher={Taylor \& Francis}
}

@article{zhao2018cross,
  title={Cross-screening in observational studies that test many hypotheses},
  author={Zhao, Qingyuan and Small, Dylan S and Rosenbaum, Paul R},
  journal={Journal of the American Statistical Association},
  volume={113},
  number={523},
  pages={1070--1084},
  year={2018},
  publisher={Taylor \& Francis}
}

@article{karmakar2019integrating,
  title={Integrating the evidence from evidence factors in observational studies},
  author={Karmakar, Bikram and French, Benjamin and Small, Dylan S},
  journal={Biometrika},
  volume={106},
  number={2},
  pages={353--367},
  year={2019},
  publisher={Oxford University Press}
}

@article{roy2022protocol,
  title={Protocol for an observational study on the effects of giving births from unintended pregnancies on later life physical and mental health},
  author={Roy, Samrat and Bogomolov, Marina and Heller, Ruth and Claridge, Amy M and Beeson, Tishra and Small, Dylan S},
  journal={arXiv preprint arXiv:2210.05169},
  year={2022}
}

@article{rosenbaum2015see,
  title={How to see more in observational studies: Some new quasi-experimental devices},
  author={Rosenbaum, Paul R},
  journal={Annual Review of Statistics and Its Application},
  volume={2},
  number={1},
  pages={21--48},
  year={2015},
  publisher={Annual Reviews}
}

@article{bogomolov2023replicability,
  title={Replicability across multiple studies},
  author={Bogomolov, Marina and Heller, Ruth},
  journal={Statistical Science},
  volume={38},
  number={4},
  pages={602--620},
  year={2023},
  publisher={Institute of Mathematical Statistics}
}

@article{ulke2021socio,
  title={Socio-political context as determinant of childhood maltreatment: a population-based study among women and men in East and West Germany},
  author={Ulke, C and Fleischer, T and Muehlan, H and Altweck, L and Hahm, S and Glaesmer, H and Fegert, J{\"o}rg M and Zenger, M and Grabe, HJ and Schmidt, S and others},
  journal={Epidemiology and Psychiatric Sciences},
  volume={30},
  pages={e72},
  year={2021},
  publisher={Cambridge University Press}
}

@article{rosenbaum2001replicating,
  title={Replicating effects and biases},
  author={Rosenbaum, Paul R},
  journal={The american statistician},
  volume={55},
  number={3},
  pages={223--227},
  year={2001},
  publisher={Taylor \& Francis}
}

@article{bekerman2024protocol,
  title={Protocol for an Observational Study on the Effects of Paternal Alcohol Use Disorder on Children's Later Life Outcomes},
  author={Bekerman, William and Bogomolov, Marina and Heller, Ruth and Spivey, Matthew and Lynch, Kevin G and Oslin, David W and Small, Dylan S},
  journal={arXiv preprint arXiv:2412.15535},
  year={2024}
}

@article{lacey2020practitioner,
  title={Practitioner review: twenty years of research with adverse childhood experience scores--advantages, disadvantages and applications to practice},
  author={Lacey, Rebecca E and Minnis, Helen},
  journal={Journal of Child Psychology and Psychiatry},
  volume={61},
  number={2},
  pages={116--130},
  year={2020},
  publisher={Wiley Online Library}
}

@article{baldwin2021population,
  title={Population vs individual prediction of poor health from results of adverse childhood experiences screening},
  author={Baldwin, Jessie R and Caspi, Avshalom and Meehan, Alan J and Ambler, Antony and Arseneault, Louise and Fisher, Helen L and Harrington, HonaLee and Matthews, Timothy and Odgers, Candice L and Poulton, Richie and others},
  journal={JAMA pediatrics},
  volume={175},
  number={4},
  pages={385--393},
  year={2021},
  publisher={American Medical Association}
}

@article{meehan2022poor,
  title={Poor individual risk classification from adverse childhood experiences screening},
  author={Meehan, Alan J and Baldwin, Jessie R and Lewis, Stephanie J and MacLeod, Jelena G and Danese, Andrea},
  journal={American journal of preventive medicine},
  volume={62},
  number={3},
  pages={427--432},
  year={2022},
  publisher={Elsevier}
}

@article{austin2024screening,
  title={Screening for adverse childhood experiences: A critical appraisal},
  author={Austin, Anna E and Anderson, Kayla N and Goodson, Marissa and Niolon, Phyllis Holditch and Swedo, Elizabeth A and Terranella, Andrew and Bacon, Sarah},
  journal={Pediatrics},
  volume={154},
  number={6},
  pages={e2024067307},
  year={2024},
  publisher={American Academy of Pediatrics Itasca, IL, USA}
}

@article{muller2025isotonic,
  title={Isotonic subgroup selection},
  author={M{\"u}ller, Manuel M and Reeve, Henry WJ and Cannings, Timothy I and Samworth, Richard J},
  journal={Journal of the Royal Statistical Society Series B: Statistical Methodology},
  volume={87},
  number={1},
  pages={132--156},
  year={2025},
  publisher={Oxford University Press UK}
}

@article{zhang2025protocol,
  title={Protocol For An Observational Study On The Effects Of Combinations Of Adverse Childhood Experiences On Adult Depression},
  author={Zhang, Ruizhe and Kong, Jooyoung and Small, Dylan S and Bekerman, William},
  journal={arXiv preprint arXiv:2502.17679},
  year={2025}
}

\orignewpage
\appendix

\section{Evidence-guided parent selection: full specification and examples}
\label{app:evidence_guided_details}
We work with the partial order on ACE combinations and the full DAG $G=(I,E)$. We form a polyforest $F$ by giving each node at most one parent. The polyforest reduces complexity but still preserves logical consistency. The in-degree in $F$ is at most one.

We first screen hypotheses and obtain a candidate index set $\mathcal{S}_\kappa$. The set $\mathcal{S}_\kappa$ contains combinations with $p$-values at most $\kappa=0.025$. We then construct a polyforest on these candidates only. For each $i\in\mathcal{S}_\kappa$, define the cover set $\mathrm{Cover}(i)\;=\;\bigl\{\,j\in\mathcal{S}_\kappa:\ X_j\succ X_i\ \text{and there is no}\ k\in\mathcal{S}_\kappa\ \text{with}\ X_j\succ X_k\succ X_i\,\bigr\}$. We choose at most one parent from this set. We use evidence to guide the choice. Specifically,
$\mathrm{par}(i)\ \in\ \arg\min_{\,j\in \mathrm{Cover}(i)}\ p_{\tau,j}^{\text{(screen)}}$, with uniform random tie-breaking. If $\mathrm{Cover}(i)=\emptyset$, then $i$ a root. This yields a polyforest $F_\kappa^{\text{val}}$ on $\mathcal{S}_\kappa$. We then compute validation-side $p$-values only for $i\in\mathcal{S}_\kappa$ and run the ISS DAG-testing algorithm on $F_\kappa^{\text{val}}$.

ISS allocates the global level $\alpha$ to roots. The procedure reallocates unused $\alpha$ to remaining roots after each rejection round. Rejection propagates along the structure. A node with strong evidence may be blocked if its parent has a large $p$-value and receives limited $\alpha$. In that case, the chain cannot unlock budget, and the strong descendant is not reached. Evidence-guided parenting reduces this risk. When constructing the polyforest, we choose each node’s parent as the candidate with the strongest available evidence, that is, the one with the smallest screening p-value. This concentrates $\alpha$ on more promising paths and therefore lowers the chance of blocking while the FWER control still holds.

\begin{figure}[H]
    \centering
    \includegraphics[width=1\linewidth]{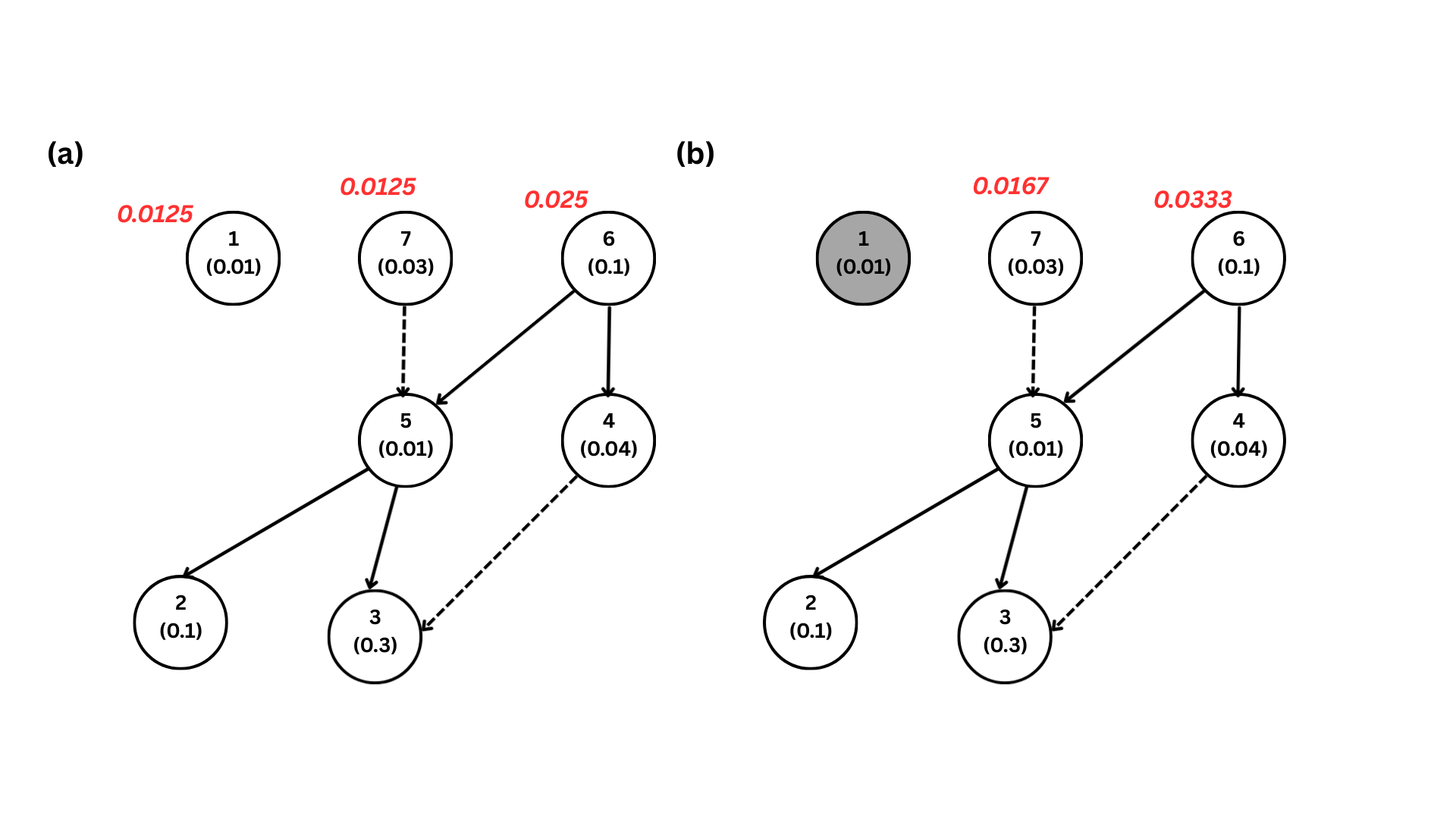}
    \caption{DAG Testing: Example 1. Each node is labeled with its $p$-value (round brackets); in the induced polyforest-weighted DAG, solid arrows indicate edges that are retained in the polyforest (weight$=$1), while dashed arrows indicate edges that exist in the full DAG but are not retained in the polyforest (weight$=$0, since each node can have at most one parent); filled circles indicate hypotheses previously rejected. Panels (a)–(b) correspond to successive iterations of the algorithm, showing how rejections propagate through the structure.}
    \label{fig:dag eg1}
\end{figure}

Figure~\ref{fig:dag eg1} shows a blocking case. Nodes $1$, $6$, and $7$ are roots and receive the initial budget. Node $5$ has a small $p$-value ($0.01$) but is attached under node $6$ with a larger $p$-value ($0.10$). Node $6$ is unlikely to be rejected with its assigned budget, so the signal at node $5$ is not reached. This is an $\alpha$-blocking event. 

\begin{figure}[H]
    \centering
    \includegraphics[width=1\linewidth]{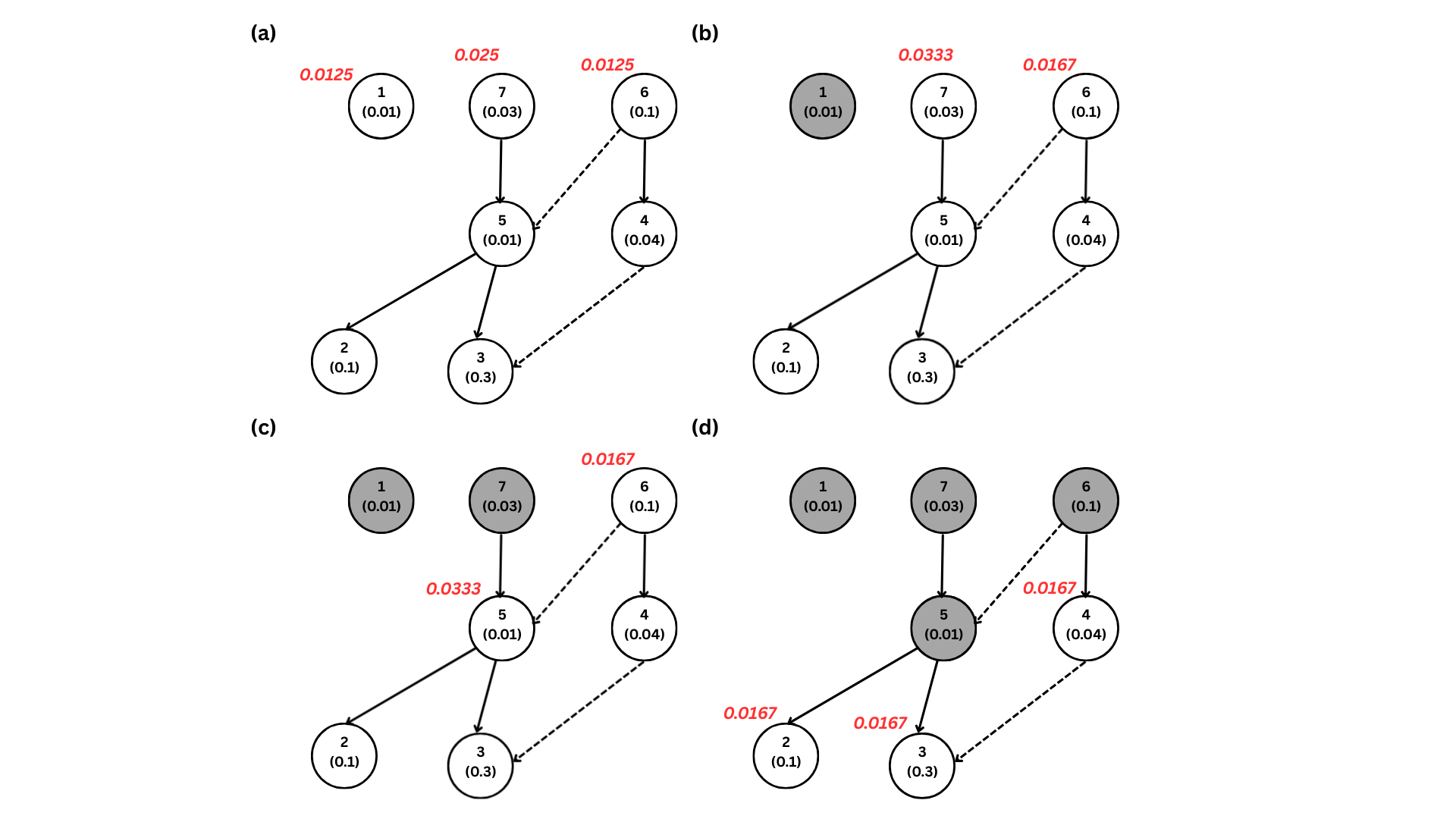}
    \caption{DAG Testing: Example 2. The setting is the same as in Figure \ref{fig:dag eg1}, except that node $5$ is attached through node $7$ rather than node $6$. Panels (a)–(d) correspond to successive iterations of the algorithm, showing how rejections propagate through the structure.}
    \label{fig:dag eg2}
\end{figure}

Figure~\ref{fig:dag eg2} shows the same graph with a different parent for node $5$. The parent is node $7$ with a smaller $p$-value ($0.03$). After node $1$ is rejected, the re-allocation increases node $7$'s budget. Node $7$ is then rejected, and node $5$ becomes reachable and is rejected as well. The evidence-guided choice avoids the bottleneck and improves discovery.

\section{Tiered $\alpha$ allocation: motivation, design and results}
\label{app:tiered allocation}

We try to optimize $\alpha$ budget allocation informed by our exploration on the blue part. One of the authors raised the question regarding Corner 3 in Table~\ref{tab:frequency coding}: Why is it that in the selected corners, \texttt{ACEDIVRC}$=1$ appears but \texttt{ACEPRISN}$=1$ does not? Intuitively, one might expect parental incarceration (\texttt{ACEPRISN}) to carry a heavier psychological burden than parental divorce (\texttt{ACEDIVRC}). The American Sociological Association’s website asserts that parental incarceration can exert more severe consequences than divorce of a parent  for both ADD/ADHD and behavioral problems.   Yet in our data-informed ISS corner selection, the block requiring \texttt{ACEDIVRC}=1 is present whereas \texttt{ACEPRISN}=1 is absent. One plausible explanation is alpha allocation in the DAG testing. The p-value for Corner 7 in Table~\ref{tab:frequency coding} is $9.26\times 10^{-7}$, while the p-value for the combination (\texttt{ACEPRISN}$=1$,\texttt{ACEHURT1}$=1$,\texttt{ACESEX}$=2$,\texttt{ACESWEAR}$=2$,\texttt{ACEADSAF}$=1$, other ACE variables $=0$) is $1.67\times10^{-5}$. Both p-values are very small, but the latter is still larger than the former. As a result, even if the two hypotheses received the same share of the $\alpha$ budget in the DAG testing procedure, it is possible for the first to be rejected while the second is not. This suggests that we may consider allocating slightly more $\alpha$ budget to combinations that are logically or theoretically more plausible, while still controlling the FWER. Therefore, we seek to combine insights from the blue data with domain knowledge to establish ordering rules that align more closely with theoretical expectations and real‐world understanding.

\begin{figure}
    \centering    \includegraphics[width=1\linewidth]{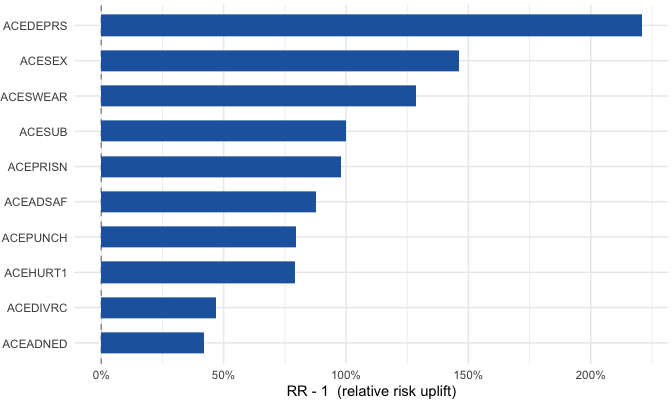}
    \caption{Marginal Ranking by Risk Ratio (RR)}
    \label{fig:marginal}
\end{figure}

\begin{figure}
    \centering    \includegraphics[width=1\linewidth]{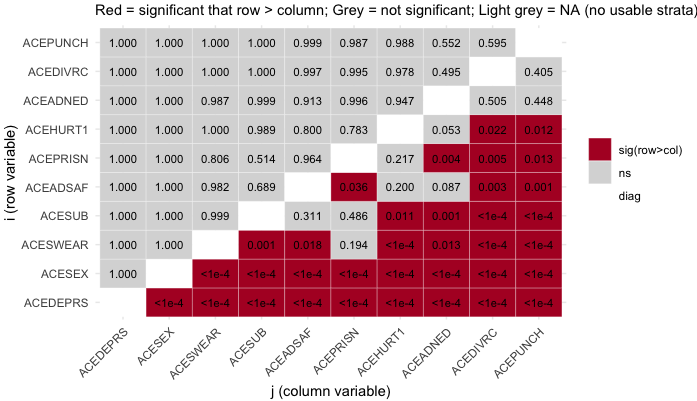}
    \caption{Conditional Dominance Heatmap}
    \label{fig:conditional}
\end{figure}

First, a marginal ranking was obtained by the risk ratio (RR) for $Y{=}1$ comparing $X_j{=}1$ vs.\ $X_j{=}0$ (Figure~\ref{fig:marginal}). This is easy to interpret but may be distorted by confounders and the co-occurence of other ACEs. 
Second, a conditional dominance map was constructed by exact stratification on the remaining ACEs and Mantel--Haenszel one-sided tests on discordant strata 1(Figure~\ref{fig:conditional}). This assesses whether, holding other ACEs fixed, the profile with $X_i{=}1, X_j{=}0$ is more strongly associated with depression than the profile with $X_i{=}0, X_j{=}1$. This controls for other ACEs but can be unstable in sparse strata and depends on the homogeneity of stratum-specific associations. We interpret significant findings as evidence of stronger conditional association. We used these two views as supporting evidence but not used as decision rules. We combined them with domain knowledge to guide our interpretation. One sociologist in our research team noted that the outcome is depression, which is a mental illness. Emotional and sexual victimization may have a stronger effect than single physical events. Repeated verbal abuse often reflects long-term hostility and rejection. Emotional neglect may be more directly linked to depression than physical neglect. Parental depression in the household acts as both a psychosocial stressor and an indicator of genetic and environmental risk. Combining these considerations yields the ordering we adopt for assigning $\alpha$-budgets: $\{\texttt{ACEDEPRS} , \texttt{ACESEX} , \texttt{ACESWEAR}\} \;>\; 
\{\texttt{ACESUB},\texttt{ACEPRISN},\texttt{ACEADSAF},\texttt{ACEHURT1}\} \;>\; 
\{\texttt{ACEADNED},\texttt{ACEDIVRC},\texttt{ACEPUNCH}\}$

The tiered $\alpha$-allocation scheme reflects both logical relations among nodes and domain priorities. At any stage of ISS, some hypotheses are already rejected, while others remain as candidates. We group the remaining nodes into subsets. Nodes in the same subset are mutually incomparable under the partial order. Across subsets, we define a simple hierarchy based on our domain-informed ordering rule. Each remaining node keeps its $\alpha$ budget from the DAG testing scheme. We then sum the $\alpha$ budgets of all candidates in a subset to form the subset’s joint $\alpha$ budget. Inside each subset, we perform a parallel gatekeeping procedure (Burman et al., 2009) and the testing follows the item-level ranking. Nodes with higher-tier ACEs (for example, \texttt{ACEDEPRS} or \texttt{ACESEX}) are tested first. If several nodes share the same priority and are logically equivalent (for example, all include \texttt{ACEDEPRS}$=1$ but differ by \texttt{ACESUB}, \texttt{ACEPRISN}, or \texttt{ACEHURT1}$=1$), we split the subset’s $\alpha$ equally among them. If no node in the current priority level is rejected, testing stops for that branch. If at least one node is rejected, we pass the remaining $\alpha$ budget from that level to the next one, dividing it according to the tier weights. This rule allows $\alpha$ budget to flow forward when evidence exists while still controlling the overall FWER.

Though, tiering may raise or lower power. The net effect is uncertain. If the data align with the domain ranking, tiering concentrates \(\alpha\) on promising chains and can increase power. If the ranking is misspecified, tiering can misallocate \(\alpha\), slow reallocation, and introduce bottlenecks in the DAG testing procedure. We therefore run a focused simulation to compare parent–selection and tiering strategies under frequency coding. The data–generating process (DGP), calibration to the target superlevel mass, and the testing machinery follow the Simulations Section, so here we only record what differs.

\begin{figure}[H]
\centering
\includegraphics[width=1\linewidth]{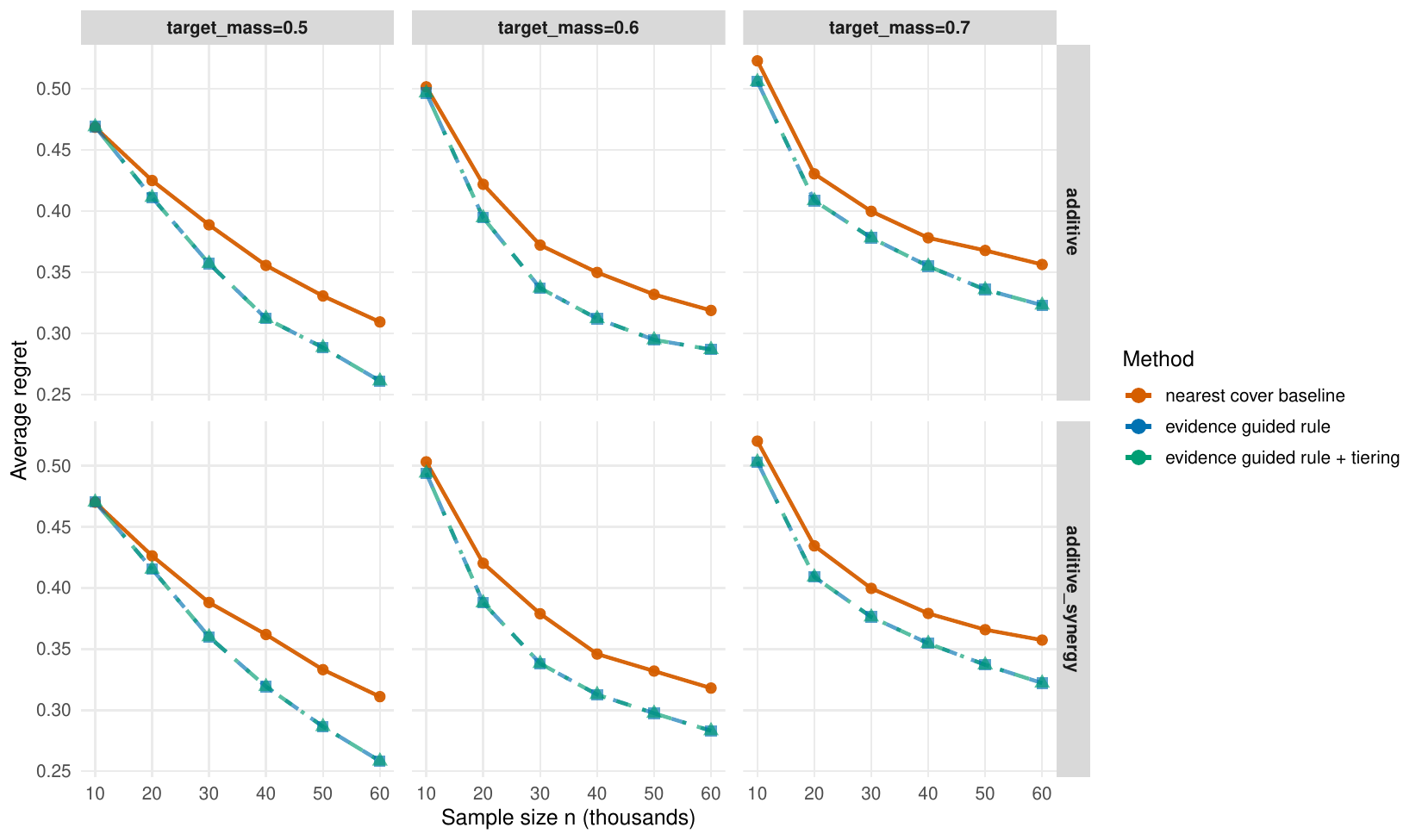}
\caption{Average regret versus sample size $n$ for the higher-risk group identified in the red part guided by the blue part. Layout and line styles match Figure~\ref{fig:union}. Notably, the blue and green curves overlap throughout.}
\label{fig:tiering}
\end{figure}

Each profile $x=(x_1,\dots,x_{10})$ has $4$ binary, $4$ three–level, and $2$ five–level coordinates, so the evaluation grid has $2^4\!\times3^4\!\times5^2=32{,}400$ cells. We split observations into blue (45\%) and red (55\%) as before and analyze the Blue$\rightarrow$Red direction only. Here, we compare four different ways. (i) Nearest–cover baseline (based on the minimal$\ell_{\infty}$ distance), (ii) evidence-guided rule, (iii) evidence-guided rule + tiering. We construct a three–tier ranking for each simulated dataset using marginal associations computed after binarizing the ACE variables. For this step, every frequency–coded item is collapsed to \(1\{x_j \ge 1\}\) vs.\ \(0\). This binarization is used only to form tiers, ISS is still run on the original frequency coding. For each configuration we still focus on the average regret, as well as  the empirical FWER as a finite sample check. The result is showcased in Figure~\ref{fig:tiering}. The evidence–guided rule (blue) and evidence guided rule + tiering (green) uniformly improves on the nearest–cover baseline (orange). Nevertheless, adding tiering on top of evidence guidance (green) yields little additional benefit: the tiered and non–tiered curves are essentially indistinguishable at the plotted scale. As expected, empirical FWER estimates remain at or below the nominal $0.05$ level in all cells. Overall, these patterns indicate that the utility of tiering is data–dependent. It helps when the domain ranking aligns with the underlying signal.

Thus, we compare ISS on the blue part with and without tiering. The universe is the frequency–coded grid \(\mathcal{U}\) with \(|\mathcal{U}|=32{,}400\). We obtain $|\widehat A^{\text{ISS}}|=7{,}276,
|\widehat A^{\text{tier}}|=6{,}252,
|\widehat A^{\text{ISS}}\cap \widehat A^{\text{tier}}|=5{,}775$. Thus \(|\widehat A^{\text{ISS}}\setminus \widehat A^{\text{tier}}|=1{,}501\) and \(|\widehat A^{\text{tier}}\setminus \widehat A^{\text{ISS}}|=477\).  Tiering selects a smaller set. It mostly trims a subset of the non–tiered discoveries and adds a modest number of alternatives. But here in the exploration, we do not use cross–split screening and also do not use evidence–guided parent selection. Both runs use the same induced polyforest (nearest–cover, $\ell_\infty$).

\orignewpage

\end{document}